\documentclass[twocolumn]{aastex63}

\usepackage{amsmath}
\usepackage{natbib}
\usepackage{multirow} 
\usepackage[utf8]{inputenc}
\usepackage[normalem]{ulem}
\useunder{\uline}{\ul}{}
\usepackage{hyperref}
\usepackage{lineno}

\hypersetup{linkcolor=cyan,citecolor=magenta,filecolor=yellow,urlcolor=blue}

\hypersetup{linkcolor=magenta, citecolor=cyan, filecolor=yellow, urlcolor=blue}

\received{ddmmyyyy}
\revised{\today}
\accepted{ddmmyyyy}
\published{ddmmyyyy}

\shorttitle{A Cosmological Fireball with Sixteen-Percent Gamma-Ray Radiative Efficiency}
\shortauthors{Li et al.}

\begin{document}

\title{A Cosmological Fireball with Sixteen-Percent Gamma-Ray Radiative Efficiency}

\author[0000-0002-1343-3089]{Liang~Li}
\affiliation{ICRANet, Piazza della Repubblica 10, 65122 Pescara, Italy}
\affiliation{INAF -- Osservatorio Astronomico d'Abruzzo, Via M. Maggini snc, I-64100, Teramo, Italy}
\affiliation{Dip. di Fisica and ICRA, Sapienza Universita di Roma, Piazzale Aldo Moro 5, I-00185 Rome, Italy}

\author{Yu~Wang}
\affiliation{ICRANet, Piazza della Repubblica 10, 65122 Pescara, Italy}
\affiliation{INAF -- Osservatorio Astronomico d'Abruzzo, Via M. Maggini snc, I-64100, Teramo, Italy}
\affiliation{Dip. di Fisica and ICRA, Sapienza Universita di Roma, Piazzale Aldo Moro 5, I-00185 Rome, Italy}

\author{Felix~Ryde}
\affiliation{Department of Physics, KTH Royal Institute of Technology, and the Oskar Klein Centre for Cosmoparticle Physics, 10691 Stockholm, Sweden}

\author{Asaf~Pe{\textquoteright}er}
\affiliation{Department of Physics, Bar-Ilan University, Ramat-Gan 52900, Israel}

\author{Bing~Zhang}
\affiliation{Department of Physics and Astronomy, University of Nevada, Las Vegas, NV 89154, USA}

\author{Sylvain~Guiriec}
\affiliation{Department of Physics, The George Washington University, 725 21st Street NW, Washington, DC 20052, USA}
\affiliation{NASA Goddard Space Flight Center, Greenbelt, MD 20771, USA}

\author{Alberto~J.~Castro-Tirado}
\affiliation{Instituto de Astrof\'isica de Andaluc\'ia (IAA-CSIC), PO Box 03004, 18008 Granada, Spain}
\affiliation{Departamento de Ingeniería de Sistemas y Automática, Escuela de Ingenierías, Universidad de Málaga, Málaga, Spain}

\author{D.~Alexander~Kann}
\affiliation{Instituto de Astrof\'isica de Andaluc\'ia (IAA-CSIC), PO Box 03004, 18008 Granada, Spain}

\author{Magnus~Axelsson}
\affiliation{Department of Astronomy, Stockholm University, SE-106 91 Stockholm, Sweden}

\author{Kim~Page}
\affiliation{School of Physics and Astronomy, University of Leicester, University Road, Leicester LE1 7RH, UK}

\author{P\'eter~Veres}
\affiliation{Center for Space Plasma and Aeronomic Research, University of Alabama in Huntsville, Huntsville,AL,USA}
\affiliation{Space Science Department,University of Alabama in Huntsville, Huntsville, AL,USA}

\author{P.~N.~Bhat}
\affiliation{Center for Space Plasma and Aeronomic Research, University of Alabama in Huntsville, Huntsville,AL,USA}

\correspondingauthor{Liang Li, Yu Wang, Bing Zhang}
\email{liang.li@icranet.org; yu.wang@uniroma1.it; zhang@physics.unlv.edu}
 
\begin{abstract}

Gamma-ray bursts (GRBs) are the most powerful explosions in the universe. How efficiently the jet converts its energy to radiation is a long-standing problem and it is poorly constrained. The standard model invokes a relativistic fireball with a bright photosphere emission component. A definitive diagnosis of GRB radiation components and measurement of GRB radiative efficiency require prompt emission and afterglow data with high-resolution and wide-band coverage in time and energy. Here we report a comprehensive temporal and spectral analysis of the TeV-emitting bright GRB 190114C. Its fluence is one of the highest of all GRBs detected so far, which allows us to perform a high-resolution study of the prompt emission spectral properties and their temporal evolution down to a timescale of about 0.1 s. We observe that each of the initial pulses has a thermal component contributing $\sim20\%$ of the total energy, the corresponding temperature and the inferred Lorentz factor of the photosphere evolve following broken power-law shapes. From the observation of the non-thermal spectra and the light-curve, the onset of afterglow corresponding to the deceleration of the fireball is considered at $\sim 6$~s. By incorporating the thermal and the non-thermal observations, as well as the photosphere and the synchrotron radiative mechanisms, we can directly derive the fireball energy budget with little dependence on hypothetical parameters and to measure a $\sim 16\%$ radiative efficiency for this GRB. With the fireball energy budget derived, the afterglow microphysics parameters can also be constrained directly from the data. 

\end{abstract}


\section{Introduction} \label{sec:intro}

On 14 January 2019 at 20:57:02.63 Universal Time (UT) (hereafter $T_{0}$), an ultra-bright burst, GRB 190114C, first triggered the Gamma-ray Burst Monitor (GBM) onboard the \emph{Fermi} Gamma-ray Space Telescope\citep{Hamburg2019} and the \textit{Neil Gehrels Swift Observatory's} (\textit{Swift} hereafter) Burst Alert Telescope (BAT)\citep{J.D.Gropp2019}. Soon after, the Large Area Telescope (LAT) onboard \emph{Fermi}, Konus-\emph{Wind}, \emph{INTEGRAL}/SPI-ACS, \emph{AGILE}/MCAL, and the \emph{Insight-HXMT}/HE were also triggered. Long-lasting and multi-wavelength afterglow observations were carried out by the Major Atmospheric Gamma Imaging Cherenkov (MAGIC) telescopes\citep{MAGICCollaboration2019} in the Teraelectronvolt band and \emph{Swift} in the X-ray and optical bands, and by several ground-based optical and radio telescopes, such as GROND\citep{Bolmer2019}, GTC\citep{Castro-Tirado2019}, VLA\citep{Alexander2019}, MeerKAT\citep{Tremou2019}.

In this paper, we present a comprehensive analysis of GRB 190114C and derive fireball parameters using rich observation data. The paper is organized as follows. In Section \ref{sec:obs}, we outline the main observing properties of the burst and the purpose of the paper. The methodology is presented in Section \ref{sec:data}. The models that apply to GRB 190114C are presented in Section \ref{sec:model}. We presented our results in Section \ref{sec:results}. Our conclusions are summarized in Section \ref{sec:conclusion}. Throughout the paper, the standard $\Lambda$-CDM cosmology with the parameters $H_{0}= 67.4$ ${\rm km s^{-1}}$ ${\rm Mpc^{-1}}$, $\Omega_{M}=0.315$, and $\Omega_{\Lambda}=0.685$ are adopted \citep{PlanckCollaboration2018}.

\section{Overview}\label{sec:obs}

The $T_{90}$ duration\footnote{The time taken to accumulate $90\%$ of the burst fluence starts at the 5\% fluence level and ends at the 95\% fluence level.} reported by the {\it Fermi} term is $\sim$116 s, and therefore GRB 190114C belongs to the class of long-duration bursts. The 1024 ms peak flux and the fluence during $T_{90}$ duration at 10-1000 keV measured by {\it Fermi}-GBM are 246.9$\pm$0.9 photon cm$^{-2}$ s$^{-1}$ and (4.436$\pm$0.005)$\times$10$^{-4}$ erg cm$^{-2}$, respectively. A measured redshift of $z$=0.424 was announced by \cite{Castro-Tirado2019}, therefore, the isotropic-equivalent $\gamma$-ray energy with a $k$-correction to the rest-frame (1-10$^{4}$ keV) is estimated to be $E_{\gamma,\rm iso}$=(2.8$\pm$0.3)$\times$10$^{53}$ erg. {\it Fermi}-LAT observed the first GeV photon at $T_0$+2.1 s, and the highest-energy photon is a 22.9 GeV event was observed at $\sim T_{0}$+15 s \citep{Wang2019}. The afterglow emission measured by {\it Swift}-XRT begins at $\sim$ $T_{0}$+68 s.

\begin{figure*}[!t]
\centering
\includegraphics[scale=1.]{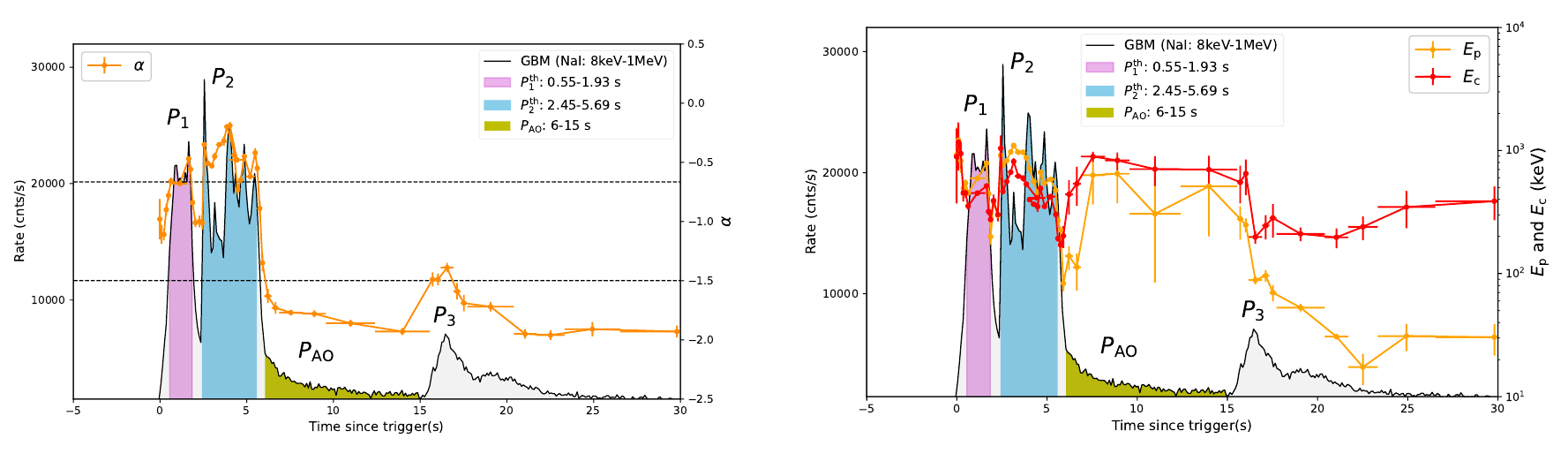}
\caption{Count lightcurve of \textit{Fermi}-GBM during the time span of $0-30$~s. Differently shaded regions marked with different colors denote the two independent thermally-sub-dominated episodes: $P^{\rm th}_{1}$ (pink) and $P^{\rm th}_{2}$ (blue), the afterglow emission episode $P_{\rm AO}$ (yellow), and the $\gamma$-ray flare emission $P_{3}$ (grey). {\bf Left panel:} Two horizontal dashed lines represent the limiting values of $\alpha$=-2/3 and $\alpha$=-3/2 for electrons in the synchrotron slow- and fast-cooling regimes, respectively. The data points connected by solid lines in orange colour represent the temporal evolution of the low-energy photon index $\alpha$ of the CPL-only fits. {\bf Right panel:}  the data points connected by solid lines in orange colour represent the temporal evolution of the $E_{\rm p}$ of the Band-only fits while those in red colour indicate the temporal evolution of $E_{\rm c}$ of the CPL-only fits.}
\label{fig:190114873_lc}
\end{figure*}

\begin{figure*}[!ht]
\centering
\includegraphics[width=1.0\hsize,clip]{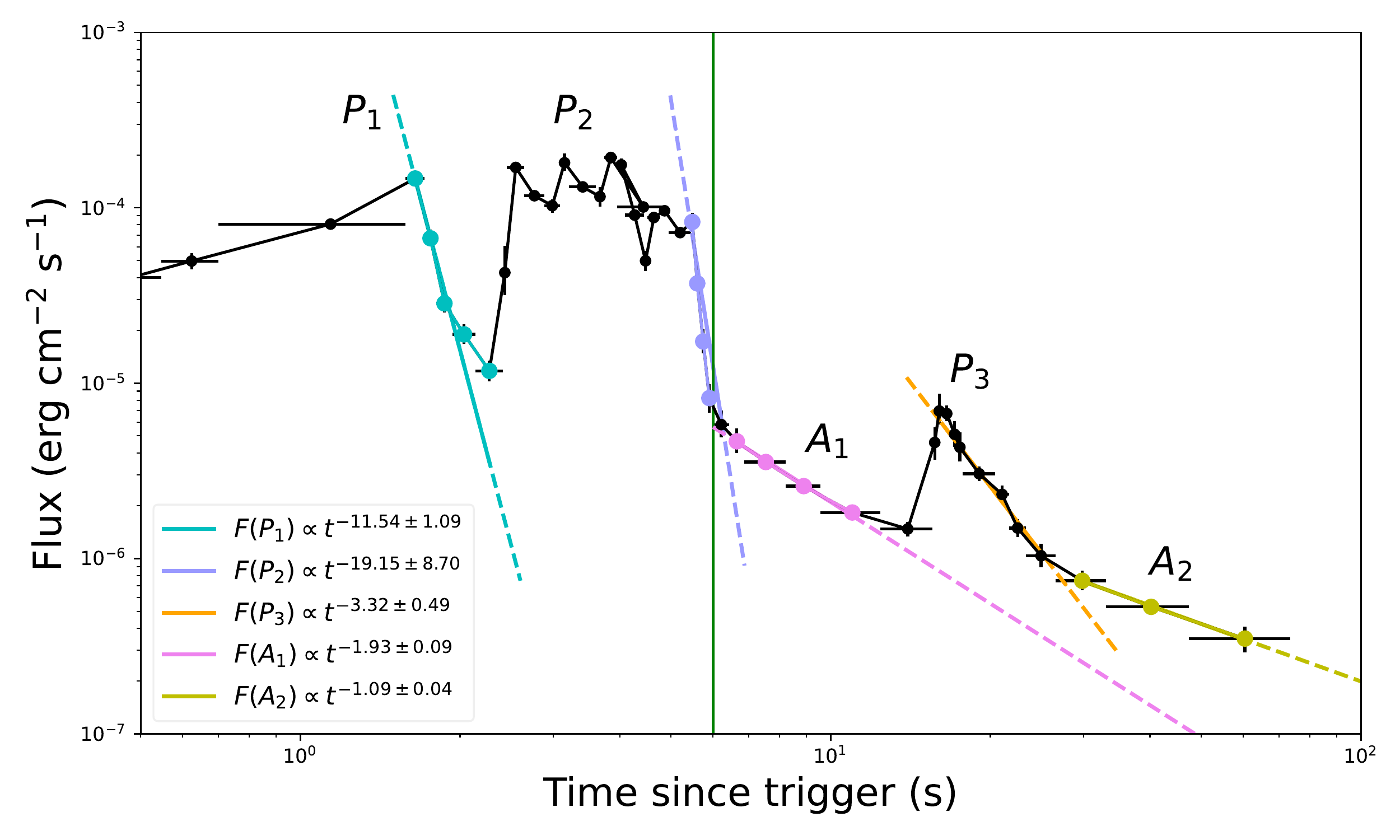} 
\caption{The GBM light curve (black dotted line) with the best fits to the decay phase of $P_{1}$ (cyan line), $P_{2}$ (blue magenta line), $P_{3}$ (orange line), and the afterglow emission $A_{1}$ (violet line) and $A_{2}$ (yellow line) using the single power-law model. Note that $A_{1}$ and $A_{2}$ correspond to the afterglow emission generated by $P_{1}$ and $P_{2}$, respectively. The onset of the afterglow emission $t_{\rm Afterglow}$ (the vertical green line) is used to estimate $\Gamma$ in equation (\ref{eq:Gamma0}). The decay index of the afterglow emission from $P_{1}$ is $\hat{\alpha}(A_{1})=-1.93\pm0.09$, which is significantly steeper than a typical value for afterglow emission measured from other GRBs, this is because part of the energy flux in this segment is clearly contributed from $P_{2}$, whereas the decay index of the afterglow emission from $P_{2}$ is $\hat{\alpha}(A_{2})=-1.09\pm0.04$, which is in good agreement with typical values found for afterglow emission.}\label{fig:p3_peak}
\end{figure*}

The prompt emission light curve consists of three distinct emission pulses \citep{ajello2020,Wang2019} (Figure \ref{fig:190114873_lc}). The first pulse (i.e., $P_{1}$) starts at $\sim$ $T_{0}$ and lasts for $\sim$ 2.35~s, the second pulse (i.e., $P_{2}$) exhibits multiple peaks and lasts from $\sim$ $T_{0}$+2.35~s to $\sim$ $T_{0}$+6~s, and the significantly fainter third pulse (i.e., $P_{3}$) extends from $\sim$ $T_{0}$+15~s to $\sim$ $T_{0}$+25~s. A majority of the $\alpha$ indices in $P_1$ and $P_2$ are beyond the line-of-death of synchrotron emission \citep[-2/3,][]{Preece1998}, suggesting the origin of the photosphere emission. Such a feature was observed in GRB 190114C and it is fully consistent with what the fireball photosphere model predicts. In the standard GRB fireball “internal-external” shock model\citep{Rees1994,Meszaros2000}, after a relativistic jet is launched from a central engine, the energy can be dissipated either by early, short-lived prompt emissions (generated by photosphere emissions where the jet becomes transparent or an internal shock via synchrotron emission), occurring at a close distance from the progenitor and mostly observed in $\gamma$-rays; or later, long-lasting, multi-wavelength afterglow emissions (generated by an external shock at a large distance where the GRB jets interact with the ambient medium), observed in X-ray, optical, and radio wavelengths. Therefore, a bright thermal component originating from the fireball photosphere and a nonthermal component presumably originating from internal shocks whose radii are greater than the photosphere radius would be expected to be observed in their prompt emission spectra. The emission in $P_3$ can be interpreted as a flare of synchrotron radiation. However, the flux, energy band and power-law decay index (Figure \ref{fig:p3_peak}) during the epoch from $\sim$ $T_{0}$+6~s to $\sim$ $T_{0}$+15~s (i.e., $P_{\rm AO}$) between $P_2$ and $P_3$ are consistent with external shock emission at $t_{\rm onset}\approx6$~s defined by the deceleration of the fireball, where $P_{\rm AO}$ represents the initialization of the afterglow emission phase generated by the deceleration of the fireball and $t_{\rm onset}$ represents the onset time of the deceleration of the fireball. The proposal that $P_{\rm AO}$ is related to afterglow emission is supported by several independent studies in the literature \citep[e.g.,][]{MAGICCollaboration2019,2019A&A...626A..12R,ajello2020, Ursi2020}. The onset signatures of afterglow emission observed in the MeV energy range during the prompt emission phase are rare, since they are typically observed as deceleration bumps in the early afterglow light curve \citep[e.g.,][]{Liang2010,Liang2013}. The afterglow phase of this GRB has the most comprehensive observations in terms of spectral coverage (see Figure \ref{fig:multi_lc_spectrum}), from radio to TeV gamma-rays\citep{MAGICCollaboration2019,2019A&A...626A..12R}. This provides a unique opportunity to study the GRB afterglow properties within the framework of the synchrotron and synchrotron self-Compton model\citep{MAGICCollaboration2019b}. Moreover, by combining the observed properties of the thermal emission in the first two episodes and of the non-thermal emission in the third episode, GRB 190114C may be the first case providing us with a unique opportunity to dissect the energy budget of a GRB fireball.

Another interesting subject related to the GRB prompt emission mechanism, which describes how efficiently the jet converts its energy to radiation, is the radiative efficiency of a burst. GRB radiative efficiency may be defined as \citep{Lloyd-Ronning2004}
\begin{eqnarray}
    \eta_\gamma & \equiv & \frac{E_\gamma}{E_{\rm tot}} = \frac{E_\gamma}{E_\gamma+E_k} = \frac{L_\gamma}{L_{w,0}}, \label{eq:eta_gam} \end{eqnarray}
where $E_\gamma$, $E_k$ and $E_{\rm tot}$ are isotropic-equivalent gamma-ray energy, afterglow kinetic energy, and total energy, respectively, and $L_\gamma$ and $L_{w,0}$ are the isotropic-equivalent average gamma-ray luminosity and total wind luminosity at the central engine, respectively. In order to evaluate the radiative efficiency of a GRB, according to Eq.(\ref{eq:eta_gam}), one needs to know the isotropic-equivalent gamma-ray energy $E_\gamma$ and the blastwave kinetic energy $E_{\rm k}$. The $E_\gamma$ term can be measured from the spectral parameters. The $E_{\rm k}$ term, on the other hand, is usually estimated from afterglow data through modeling, but the estimated values typically have large uncertainties \citep{Zhang2007b}. By combining the prompt emission and early afterglow data, \cite{Zhang2021} proposed a new method to directly dissect the GRB fireball energy budget into three components and measure their values. As a result, GRB radiation efficiency can also be directly calculated with little uncertainty. The method requires a GRB with a dominant thermal spectral component, a deceleration bump feature in the early afterglow light curve, and a measured redshift. The measured parameters include the initial dimensionless specific enthalpy ($\eta$), bulk Lorentz factors at the photosphere radius ($\Gamma_{\rm ph}$), and before fireball deceleration ($\Gamma_{0}$), the amount of mass loading ($M$), and the GRB radiative efficiency ($\eta_\gamma$). These measured parameters only weakly depend on the density $n$ of the interstellar medium when the composition ${\cal Y}$ parameter (typically unity) is specified, where ${\cal Y}$ is the lepton-to-baryon number ratio, which equals unity for a pure hydrogen fireball but could be greater (for a pair-loaded fireball) or slightly smaller (for a neutron-rich fireball without pair loading) than unity.

Once these fireball parameters can be precisely measured, one can also estimate the blastwave kinetic energy as $E_{\rm K}=\Gamma_0 M c^{2}$, as well as the GRB radiative efficiency $\eta_\gamma$. In this paper, using the photosphere data observed in $P_1$ and $P_2$, and the early afterglow data observed in $P_{\rm AO}$ as supported by several independent studies in the literature, as well as a measured redshift, we apply the method proposed in \cite{Zhang2021} to directly dissect the GRB fireball energy budget and therefore to measure GRB radiative efficiency for GRB 190114C.

\section{Methodology} \label{sec:data}

\subsection{Data Reduction}

We reduced the GBM data using a Python package, namely, \emph{The Multi-Mission Maximum Likelihood Framework} (3ML,\citealt{Vianello2015}). The data we used for our spectral analysis includes the two most strongly illuminated sodium iodide (NaI) scintillation detectors (n3, n4) and the most-illuminated bismuth germanium oxide (BGO) scintillation detector (b0) onboard {\it Fermi}-GBM, as well as the corresponding response files (.rsp2 files are adopted). The detector selections were made by considering an angle of incidence less than\citep{Goldstein2012, NarayanaBhat2016} $60^{\circ}$ for NaI and the lowest angle of incidence for BGO. The Time-Tagged Event (TTE) data type is used for the NaI data (8 keV--1 MeV) and BGO data (200 keV--40 MeV). To avoid the K-edge at 33.17 keV, the spectral energy range was also cut from 30 to 40 keV. The background fitting is chosen using two off-source intervals, including the pre-burst (-20$\sim$-10~s) and post-burst (180$\sim$200~s) epochs, and with the polynomial order determined (0-4) by applying a likelihood ratio test. The source interval is selected over the duration (-1$\sim$116~s) reported by the {\it Fermi}-GBM team. The maximum likelihood-based statistics, the so-called Pgstat, are used, given by a Poisson (observation)-Gaussian (background) profile likelihood\citep{Cash1979}.

\subsection{Bayesian Spectral Analysis} 

The spectral parameters are obtained by adopting a fully Bayesian analysis approach. The main idea is that after the experimental data are obtained, Bayes's theorem is applied to infer and update the probability distribution of a specific set of model parameters. Building up a Bayesian profile model ($M$), and given an observed data set ($D$), the posterior probability distribution $p(M\mid D)$, according to the Bayes's theorem, is given by 
\begin{equation}
p(M\mid D)=\frac{p(D\mid M)p(M)}{p(D)},
\end{equation}
where, $p(D\mid M$) is the likelihood that combines the model and the observed data, and expresses the probability to observe (or generate) the dataset $D$ from a given model $M$ with its parameters; $p(M)$ is prior on the model parameters; and $p(D)$ is called the evidence, which is a constant with the purpose of normalisation. We utilise the typical spectral parameters from the $Fermi$-GBM catalogue as the prior distributions:
\begin{eqnarray}
\label{n} \left\{ \begin{array}{ll}
A_{\rm Band} \sim \log  \mathcal{N} ~(\mu=0,\sigma=2) & \rm cm^{-2} \ keV^{-1} \ s^{-1}\\
\alpha_{\rm Band} \sim \mathcal{N} ~(\mu=-1,\sigma=0.5) \\
\beta_{\rm Band} \sim \mathcal{N} ~(\mu=-2,\sigma=0.5) \\
E_{\rm Band} \sim \log \mathcal{N} ~(\mu=2,\sigma=1) & \rm keV\\
A_{\rm CPL} \sim \log  \mathcal{N} ~(\mu=0,\sigma=2) & \rm cm^{-2} \ keV^{-1} \ s^{-1}\\
\alpha_{\rm CPL} \sim \mathcal{N} ~(\mu=-1,\sigma=0.5) \\
E_{\rm CPL} \sim \log \mathcal{N} ~(\mu=2,\sigma=1) & \rm keV\\
A_{\rm BB} \sim \log  \mathcal{N} ~(\mu=-4,\sigma=2) & \rm cm^{-2} \ keV^{-1} \ s^{-1}\\
kT_{\rm BB} \sim \log \mathcal{N} ~(\mu=2,\sigma=1) & \rm keV\\
\end{array} \right.
\label{eq:prior}
\end{eqnarray}
We employ a Markov Chain Monte Carlo (MCMC) sampling method ($emcee$, \citealt{Foreman-Mackey2013}) to sample the posterior. The parameter estimation is obtained at a maximum a posteriori probability from the Bayesian posterior density distribution, and its uncertainty (or the credible level) is evaluated from the Bayesian highest posterior density interval at 1$\sigma$ (68\%) Bayesian credible level.

\subsection{Model Comparison}
  
The best-fit model is reached by comparing the DIC values of different models and picking the one with the lowest value. The DIC is defined as DIC=-2log[$p$(data$\mid\hat{\theta}$)]+2$p_{\rm DIC}$, where $\hat{\theta}$ is the posterior mean of the parameters, and $p_{\rm DIC}$ is the effective number of parameters. The preferred model is the one that provides the lowest DIC score. We report the $\Delta$DIC values by comparing the best model with other models in Table \ref{tab:model}. Log(posterior) is adopted by the method of the maximum likelihood ratio test, which is treated as a reference of the model comparison\citep{Vuong1989}.
  
\subsection{BBlocks Methods}
  
We use a method called Bayesian blocks (BBlocks, \citealt{Scargle2013}) to rebin the Time Tagged Event (TTE) light curve. Time bins are selected in such a way as to capture the true variability of the data. Such a calculation requires each bin to be consistent with a constant Poisson rate. In each bin, it allows for a variable time width and signal-to-noise (S/N) ratio. As such, we first apply the BBlocks method with the false alarm probability $p_{0}=0.01$ to the TTE light curve of the most strongly illuminated GBM detector (n4). The other detectors (n3 and b0) are then binned into matching slices. We notice that the BBlocks analysis generates two slices ($0.70 \sim 1.58$~s and $1.58 \sim 1.71$~s) from $0.70$~s to $1.71$~s. On the other hand, the two slices have a very high significance (263.97 and 115.59). To study the parameter evolution in great detail, we, therefore, rebin the time intervals with five narrower slices of significance $>$ $80$ instead. We also did the same analysis on the last slice of $P_{2}$ ($5.51 \sim5.69$~s), generating two narrower slices ($5.51\sim5.65$~s and $5.65\sim5.69$~s), with significance $>$ $70$ each, to study the temperature evolution in more detail. We, therefore, obtain 8 slices for $P^{\rm th}_{1}$ and 16 slices for $P^{\rm th}_{2}$ to study the photosphere properties.

\section{Models} \label{sec:model}

\subsection{Deriving the Photosphere Properties Using the Traditional Method}

The traditional method to derive the photosphere properties invokes the standard fireball model\citep{Meszaros2000, Peer2015,Zhang2018}. Within this framework, the fireball invokes thermally accelerated, matter-dominated, and finally shocks-decelerated ejecta\citep{Goodman1986,Paczynski1986}.

The thermal emission of GRB 190114C is extremely strong, ranking second in thermal-to-total flux ratio (21\%) among the over 2700 GRBs observed by {\it Fermi}-GBM up to date. The identification of the strong thermal component in GRB 190114C allows us to determine the physical properties of the relativistic outflow within the framework of the non-dissipative photosphere theory\citep{Peer2007,vereshchagin2017}, which also applies to moderately dissipated photospheres where the photosphere spectrum is only mildly modified. The photosphere photons observed at a given time, corresponding to the one-time bin in our time-resolved analysis, are assumed to be emitted from an independent thin shell. Therefore, the observed BB temperature $kT_{\rm obs}$, the BB flux $F_{\rm BB}$, and the total flux $F_{\rm tot}$ (thermal+non-thermal) of a given time bin determine the photosphere properties of the corresponding shell. The entire duration of photosphere emission is conjugated by the emissions from a sequence of such shells. One can infer the bulk Lorenz factor $\Gamma$, and the initial size of the flow $R_{0}$ in each time bin and their temporal evolutions.

Within the framework of the standard fireball model~\citep{Peer2007}, for a given shell, it is generated at an initial radius
\begin{equation}
r_{0}(r_{\rm ph}>r_{\rm s})=\frac{4^{3/2} d_{\rm L}}{(1.48)^{6} \xi^{4} (1+z)^{2}}(\frac{F^{\rm obs}_{\rm BB}}{\mathbb{Y} F^{\rm obs}})^{3/2} \Re,
\end{equation}
and self-accelerates to reach a saturated Lorentz factor
\begin{equation}
\eta(\equiv\Gamma)(r_{\rm ph}>r_{\rm s})= \left[\xi (1+z)^{2} d_{\rm L} \left(\frac{\mathbb{Y} F^{\rm obs}\sigma_{\rm T}}{2 m_{\rm p} c^{3} \Re}\right)\right]^{1/4}
\label{eq:Gamma}
\end{equation}
in the coasting phase. If the photosphere radius is greater than the saturation radius, it reads 
\begin{equation}
r_{\rm ph}(>r_{\rm s})=\frac{L_{0}\sigma_{T}}{8 \pi m_{\rm p} c^{3} \Gamma^{3} },
\end{equation}
where the dimensionless parameter
\begin{equation}
\Re=\left(\frac{F_{\rm BB}}{\sigma_{\rm B} T^{4}}\right)^{1/2}=\xi \frac{(1+z)^{2}}{d_{\rm L}} \frac{r_{\rm ph}}{\Gamma}
\end{equation}
presents the effective transverse size of the photosphere. The burst luminosity $L_{0}=4\pi d^{2}_{L} \mathbb{Y} F_{\rm tot}$ is given by the observation, $\mathbb{Y}$ is the ratio between the total fireball energy and the energy emitted in gamma-rays. The numerical factor $\xi$ is of the order of unity that can be obtained from angular integration. The luminosity distance $d_{\rm L}$ of redshift $z$ is integrated by assuming the standard Friedmann–Lemaître–Robertson–Walker (FLRW) metric. Other physical constants are the Thomson cross section $\sigma_{\rm T}$, the proton rest mass $m_{\rm p}$, the speed of light $c$, and the Stefan-Boltzmann constant $\sigma_{\rm B}$.

\subsection{Directly Deriving the Fireball Properties from Observations} \label{sec:newmodel}

GRB 190114C has a redshift measurement. Its prompt emission is thermally sub-dominated and its lightcurve has a clear early pulse indicating the afterglow initiation. These three properties make it the first case where one can use observational properties to directly determine the fireball characteristics including the dimensionless specific enthalpy at the engine $\eta$, isotropic equivalent total mass $M$, bulk Lorentz factor at the site of the photosphere $\Gamma_{\rm ph}$, initial afterglow Lorentz factor before the deceleration phase $\Gamma_{\rm s}$, the kinetic energy in the fireball $E_{\rm k}$, and $\gamma$-ray radiative efficiency $\eta_\gamma$. The method described below follows \cite{Zhang2021}.

The initial, total energy of a fireball is
\begin{equation}
E_{\rm tot} = \eta M c^2.
\end{equation}
The fireball undergoes rapid acceleration and reaches a Lorentz factor $\Gamma_{\rm ph}$ at the photosphere. 
The internal energy released as thermal emission can be estimated as
\begin{equation}
E_{\rm th} = (\eta-\Gamma_{\rm ph}) M c^2,
\label{eq:Eth}
\end{equation}
Afterwards, the fireball moves at an almost constant speed until internal dissipation at internal shocks occurs at a larger distance.
The emitted non-thermal emission can be estimated as
\begin{equation}
E_{\rm nth} = (\Gamma_{\rm ph}-\Gamma_0) M c^2,
\label{eq:Enth}
\end{equation}
where $\Gamma_{\rm s}$ is the Lorentz factor after the dissipation and also the initial Lorentz factor in the afterglow phase. 

The Lorentz factor at photosphere radius $\Gamma_{\rm ph}$ can be estimated as (modified from \cite{Peer2007, Begue2014}, see \cite{Zhang2021} for details)
\begin{equation}
\begin{split}
    \Gamma_{\rm ph} 
     &=  \left[(1+z)^2 D_{\rm L} \frac{ {\cal Y}\sigma_{\rm T} F^{\rm obs}_\gamma}{2 m_p c^3 {\cal R}} \frac{\eta^{3/2}}{\eta-\Gamma_0} \right]^{2/9}, \\
{\cal R}&=\left(\frac{F^{\rm obs}_{\rm BB}}{\sigma_{\rm B} T^{4}}\right)^{1/2}.
\end{split}
\label{eq:Gammaph_new}
\end{equation}
which involves several direct observables including redshift $z$, total flux $F^{\rm obs}_\gamma$, thermal flux $F^{\rm obs}_{\rm BB}$ and the observed temperature $T$. Other parameters are the pair multiplicity parameter ${\cal Y}$ which is commonly taken as $1$, the luminosity distance $D_{\rm L}$ computed from the redshift adopting the FLRW cosmology, and fundamental constants such as speed of light $c$, proton mass $m_{\rm p}$, Thomson cross section $\sigma_{\rm T}$, and Stefan-Boltzmann constant $\sigma_{\rm B}$.

The initial Lorentz factor of the afterglow phase $\Gamma_{\rm s}$ can be derived by equating the kinetic energy to the swept-up ISM mass at the deceleration time $t_{\rm dec}$, which is an observable indicated by a light-curve pulse (the third pulse for 190114C). Using Eq.(7.81) of \cite{Zhang2018} and the above arguments, we derive
\begin{equation}
    \Gamma_{\rm s}  \simeq  170 t_{\rm dec,2}^{-3/8} \left(\frac{1+z}{2}\right)^{3/8} \left(\frac{E_{\rm th,52}+E_{\rm nth,52}}{n}\right)^{1/8} \left(\frac{\Gamma_0}{\eta-\Gamma_0} \right)^{1/8}.
\label{eq:Gamma0}
\end{equation}
where $n$ is the ISM density assumed as one particle per cubic centimetre as usual\footnote{Note that we do not discuss the case of a wind medium \citep{Dai1998MNRAS,Meszaros1998,Chevalier1999} in our theoretical model \citep{Zhang2021}. This is because afterglow observations suggest that the majority of GRBs, especially those with the clear deceleration signature, are consistent with having a constant density medium \citep{Zhang2007b,Liang2010}. More importantly, because for a wind medium the fireball dynamics should be in the ``thick shell'' regime \citep{Kobayashi2003,Wu2003} while the observations do not require a thick shell dynamics for this burst, we only consider a constant-density medium in our calculation.}. 

Simultaneously solving Eqs. \ref{eq:Eth} -- \ref{eq:Gamma0}, we obtain fireball parameters $\eta$, $\Gamma_{\rm ph}$, $M$ and $\Gamma_{\rm s}$, and in turn, we can calculate the kinetic energy of the afterglow 
\begin{equation}
E_k = \Gamma_0 M c^2,
\end{equation}
and the efficiency of the prompt gamma-ray emission
\begin{equation}
\eta_\gamma = \frac{E_{\rm th}+E_{\rm nth}}{E_{\rm tot}} = \frac{\eta-\Gamma_0}{\eta}. 
\label{eq:eta_gam2}
\end{equation}

\section{Results} \label{sec:results}

\subsection{Multi-wavelength Observations}
  
\begin{figure}
\includegraphics[scale=1]{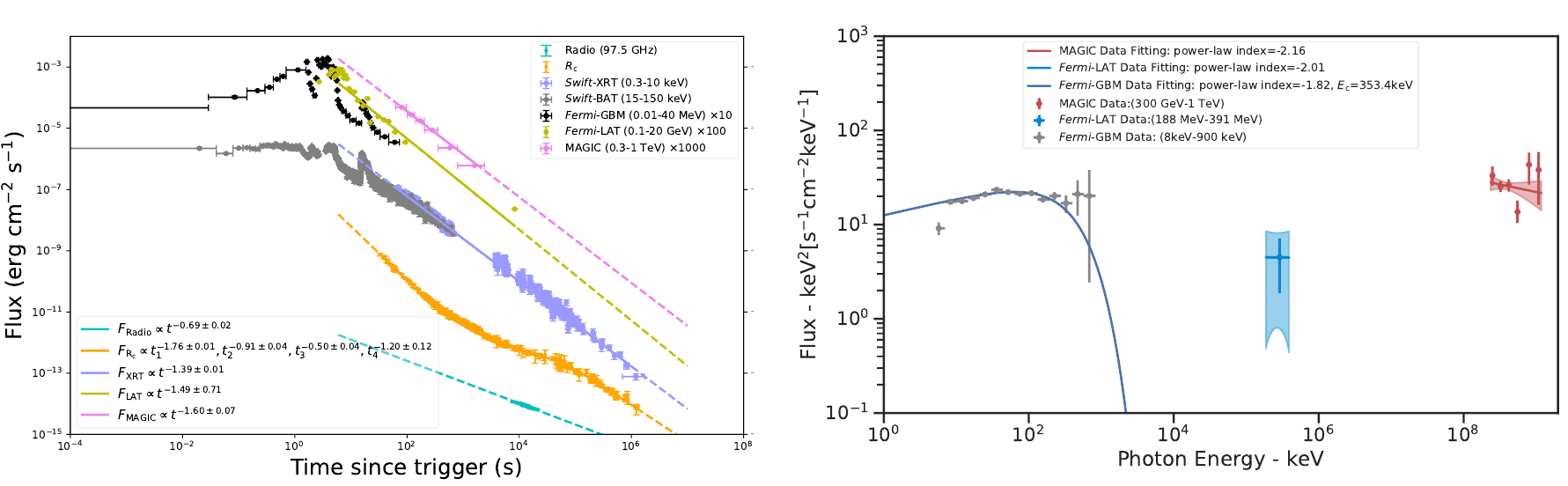}
\caption{{\bf Left panel:} multi-wavelength light curve. The data points indicated by violet, yellow, black, grey, blue magenta, orange, and cyan represent the MAGIC, {\it Fermi}-LAT, {\it Fermi}-GBM, {\it Swift}-BAT, {\it Swift}-XRT, the optical ($R_{\rm c}$-band), and the radio (at 97.5 GHz) observations, respectively. The solid lines are the best power-law fitting to the data. Note that: (1) The LAT data are separated into two parts at $\sim$6~s, and here we only fit the second (afterglow) part ($>$6~s). (2) The optical $R_{\rm c}$-band has been corrected for Galactic and host extinction, and the contribution from the host galaxy has also been subtracted. This light curve has been created by shifting data from different bands to the $R$ band. {\bf Right panel:} multi-wavelength spectrum, covering the energy in MeV (grey), GeV (blue), and TeV (jacinth) emission, which is simultaneously observed from $T_{0}$+68~s to $T_{0}$+110~s by {\it Fermi}-GBM, {\it Fermi}-LAT, and MAGIC, respectively.}
\label{fig:multi_lc_spectrum}
\end{figure}

(1) {\it TeV (MAGIC) Observations:} The Major Atmospheric Gamma Imaging Cherenkov (MAGIC) telescopes observed for the first time very-high-energy gamma-ray ($>$ 1 TeV) emission from $T_{0}$+57~s until $T_{0}$+15912~s~\citep{MAGICCollaboration2019}, setting the record of one of the highest energy photon detected from any GRB. Both the TeV lightcurve and spectrum can be well-described by a power-law model\footnote{The convention $F_{\nu,t}=t^{\hat{\alpha}}\nu^{\hat{\beta}}$ is adopted throughout the paper.}, with the temporal decay index $\hat{\alpha}_{\rm MAGIC}$=-1.60$\pm$0.07 (Figure \ref{fig:multi_lc_spectrum}a) and the spectral decay index $\hat{\beta}_{\rm MAGIC}$=-2.16$\pm$0.30 (Figure \ref{fig:multi_lc_spectrum}b). The total TeV-band (0.3-1 TeV) energy integrated between $T_{0}$+62~s and $T_{0}$+2454~s is $E^{\rm MAGIC}_{\gamma,\rm iso}$$\sim$ 4.0$\times$10$^{51}$~erg (\citealt{MAGICCollaboration2019}).

(2) {\it GeV ({\it Fermi}-LAT) Observations:} The first GeV photon was observed by {\it Fermi}-LAT at $T_{0}$+2.1~s. The highest-energy photon detected by LAT is a 22.9 GeV event detected at $T_{0}$+15~s\citep{Kocevski2019}, therefore, the bandwidth (0.1-10 GeV) was reasonably adopted to measure the total GeV energy detected by LAT. For comparison, in Table \ref{tab:energy} we also report the results based on the other two bandwidths: (0.1-100 GeV) and (0.1-1 GeV). After that time, the lightcurve and spectrum as measured by LAT (0.1-10 GeV) from $T_{0}$+55~s to $T_{0}$+8000~s are well-fitted by a power-law model with temporal decay index $\hat{\alpha}_{\rm LAT}$=-1.29$\pm$0.01 (Figure \ref{fig:multi_lc_spectrum}a) and spectral slope index $\hat{\beta}_{\rm LAT}$=-2.01$\pm$0.98 (Figure \ref{fig:multi_lc_spectrum}b). The total GeV-band (0.1-10 GeV) energy integrated between $T_{0}$+2.1~s and $T_{0}$+8000~s is $E^{\rm LAT}_{\gamma,\rm iso}$ =(1.09$\pm$0.24)$\times$10$^{53}$~erg, which can be separated into two emission components: the prompt emission ($\leq$ 6~s) accounts for $E^{\rm LAT}_{\gamma,\rm iso}$=(8.49$\pm$1.80)$\times$10$^{51}$~erg, while the afterglow emission ($>$6~s) accounts for $E^{\rm LAT}_{\gamma,\rm iso}$=(1.01$\pm$0.24)$\times$10$^{53}$~erg.

(3) {\it MeV ({\it Fermi}-GBM) Observations:} The duration of the GRB ($T_{90}$) is about $116$~s as reported by {\it Fermi}-GBM. The 1024~ms peak flux and the fluence at 10-1000 keV measured by {\it Fermi}-GBM are 246.9$\pm$0.9 photon cm$^{-2}$ s$^{-1}$ and (4.436$\pm$0.005)$\times$10$^{-4}$ erg cm$^{-2}$, respectively. With a known redshift, $z$=0.4245 $\pm$0.0005\citep{Selsing2019}, and based on the best models for each emission episode (CPL+BB model for prompt emission and Band model for afterglow emission), the total $k$-corrected isotropic energy in the rest-frame 1-10$^{4}$ keV band as derived from {\it Fermi}-GBM observations between $T_{0}$+0~s and $T_{0}$+116~s is $E^{\rm GBM}_{\gamma,\rm iso}$=(2.82$^{+0.43}_{-0.25}$)$\times$10$^{53}$~erg\citep{Wang2019}, and between $T_{0}$+15~s and $T_{0}$+25~s ($P_3$) is $E^{\rm GBM}_{\gamma,\rm iso}$=$\sim$1.24$\times$10$^{52}$~erg . The prompt emission ($\leq$ 6~s) accounts for $E^{\rm GBM}_{\gamma,\rm iso}$=(2.29$^{+0.10}_{-0.09}$) $\times$ 10$^{53}$ erg while the afterglow emission ($>$6~s) accounts for $E^{\rm GBM}_{\gamma,\rm iso}$=(5.33$^{+4.23}_{-2.34}$) $\times$ 10$^{52}$~erg. There is a $\sim$ 3.24~s lag between the GBM emission and the LAT emission. 

(4) {\it keV ({\it Swift}-XRT) Observations:} Following the trigger by {\it Swift}-BAT, the spacecraft slewed immediately to the location of the burst. The X-ray Telescope (XRT) began observing the afterglow at $T_{0}$+64~s. Pointed Windowed Timing mode data were collected from $T_{0}$+68~s to $T_{0}$+626~s, after which the count rate was low enough for Photon Counting mode to be utilised. The burst was followed for more than 28 days, although the last detection occurred on $T_{0}$+20~day. The XRT lightcurve showed a typical power-law behaviour  with a power-law index $\hat{\alpha}_{\rm XRT}$=-1.39+0.01 (Figure \ref{fig:multi_lc_spectrum}). The isotropic X-ray energy release $E^{\rm XRT}_{\rm X, iso}$ measured by {\it Swift}-XRT (0.3-10 keV) from $T_{0}$+68~s to $T_{0}$+13.86~days is $\sim$1.48$\times$10$^{52}$~erg.

(5) {\it Optical Observations:} Optical data have been gathered from Refs.\citep{MAGICCollaboration2019,Jordana-Mitjans2020,Misra2019,2021arXiv211204759M}, as well as GCN data from Refs. \citep{Bikmaev2019,Im2019,Im2019a,Kim2019a,Kim2019,Mazaeva2019,Watson2019,Watson2019a}. The automatically processed UVOT data are also used. All afterglow data have been host-subtracted using the host-galaxy values taken from \cite{deUgartePostigo2020}, and only late data without any supernova contribution has been used. Note that \cite{Jordana-Mitjans2020} found chromatic evolution in their early RINGO3 data. However, this effect is small, which leads to some additional scatter around the first and second steep-to-shallow decay transition. After the respective host galaxy magnitude has been subtracted for each band, all the bands are shifted to the $R_{\rm c}$ band to produce a composite light curve stretching from 33 s to 56.4 days after the GRB trigger. The light curve can be described by multiple power-law decay segments in a steep-normal-shallow-normal-steep arrangement. The first two segments have slopes $\hat{\alpha}_{\rm opt,1}=-1.628\pm0.012$ and $\hat{\alpha}_{\rm opt,2}=-1.035\pm0.006$, with a break time at $t_{b,1}=429\pm61$ s and a sharp transition index with $n=-13.3\pm2.0$ (Figure \ref{fig:multi_lc_spectrum}a), such an early steep-normal transition is consistent with the superposition of a reverse shock component with a forward shock component. After a second, sharp break ($n=-8.6\pm1.9$) at $t_{\rm b,2}=4856\pm216$ s, the lightcurve goes over into an even flatter phase, decaying with $\hat{\alpha}_{opt,3}=-0.512\pm0.035$. At a break time $t_{b,3}=0.548\pm0.036$ d (with a smoother transition index $n=4.4\pm2.1$), the decay becomes steeper again, reaching a value similar to $\hat{\alpha}_{\rm opt,2}$, $\hat{\alpha}_{opt,4}=-1.146\pm0.036$, indicating the shallow decay phase may be interpreted as an energy injection. We find a final break at $t_{\rm b,4}=6.33\pm1.26$ d to an even steeper decay $\hat{\alpha}_{\rm opt,5}=-1.714\pm0.041$ ($n=10$ had to be fixed).

This final break may represent a jet break. If so, the post-break slope would be quite shallow but not unprecedented (see the sample of \citealt{Zeh2006} for comparison). There is no conclusive evidence from the X-ray data for this break. However, we note the last three \emph{Swift} data points are decaying more steeply than before, and the X-ray data only extends to $\approx14$ d, which does not allow strong conclusions to be drawn.

(6) {\it Radio Observations:} The radio data points are taken from \citep{Laskar2019}. Radio observations were carried out by the Atacama Large Millimeter/submillimeter Array (ALMA) at Band 3 with a center frequency of 97.5 GHz spanning the period from $T_{0}$+0.0995 days to $T_{0}$+0.217 days and lasting for 3 hours, and together with NSF’s Karl G. Jansky Very Large Array (VLA) observations with a full sequence of observations spanning 5–38 GHz, starting at $T_{0}$+0.197 days and ending at $T_{0}$+0.261 days. As shown in the left panel of Figure \ref{fig:multi_lc_spectrum}, the radio afterglow lightcurve from the ALMA observation is well-fitted with a power-law model with the temporal decay index $\hat{\alpha}_{\rm radio}=-0.69\pm0.02$. The radio observations at $\lesssim T_{0}$+0.03 days, as well as the optical and millimetre observations, were interpreted as emissions from the reverse-shocked ejecta in \citep{Laskar2019}.

\subsection{Time-integrated and Time-resolved Spectral Analysis}

\begin{figure}
\center
\includegraphics[width=0.9\hsize,clip]{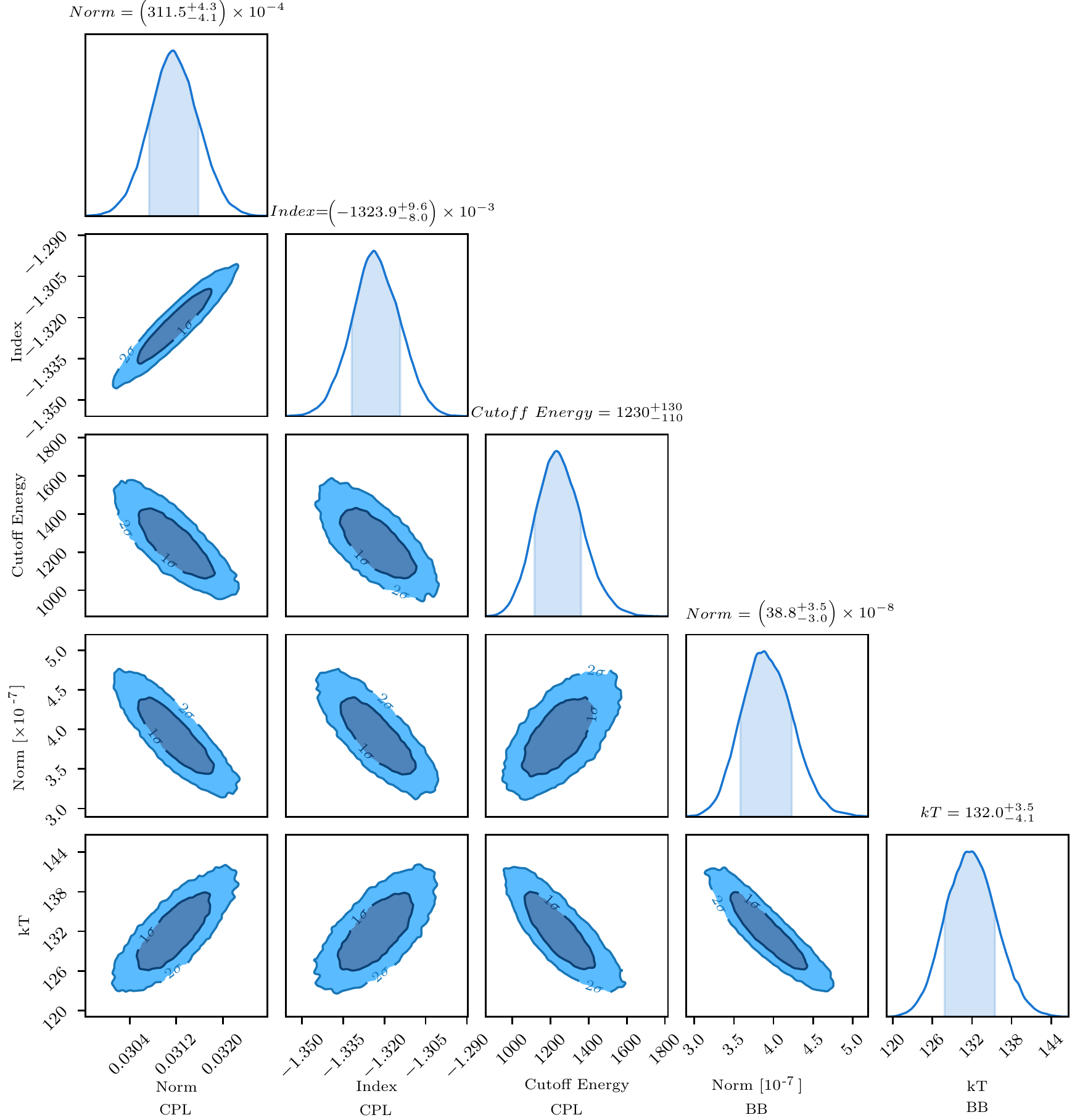}
\caption{Bayesian Monte Carlo Time-integrated spectral fits for \textit{Fermi}-GBM data from $T_{0}+0$~s to $T_{0}+116$~s ($T_{90}$ duration). We apply $20$ chains, each chain iterates $10^4$ times and burns the first $10^3$ times. The parameters are normalisation (Norm CPL), cut-off energy and power-law index of the cut-off power-law model, as well as normalisation (Norm BB) and temperature ($kT$) of the BB model.}
\label{fig:datafitting}
\end{figure}

We first performed the time-integrated spectral analysis (treating the entire $T_{90}$ as one-time bin, i.e., from $T_{0}$ to $T_{0}+116$~s) by using various GRB spectral models, including power-law (PL), blackbody (BB), cutoff power law (CPL), Band function\citep{Band1993}, smoothly broken power law (SBKL), PL+BB\citep{Ryde2005}, PL+Bandcut, CPL+BB\citep{Li2019c}, and Band+BB\citep{Guiriec2011}, respectively. Our refined time-integrated spectral analysis suggests that the CPL+BB model can best characterize the spectral shape of the burst. The corresponding corner plot is shown in Figure \ref{fig:datafitting}.

\begin{figure*}
\centering
\includegraphics[scale=0.7]{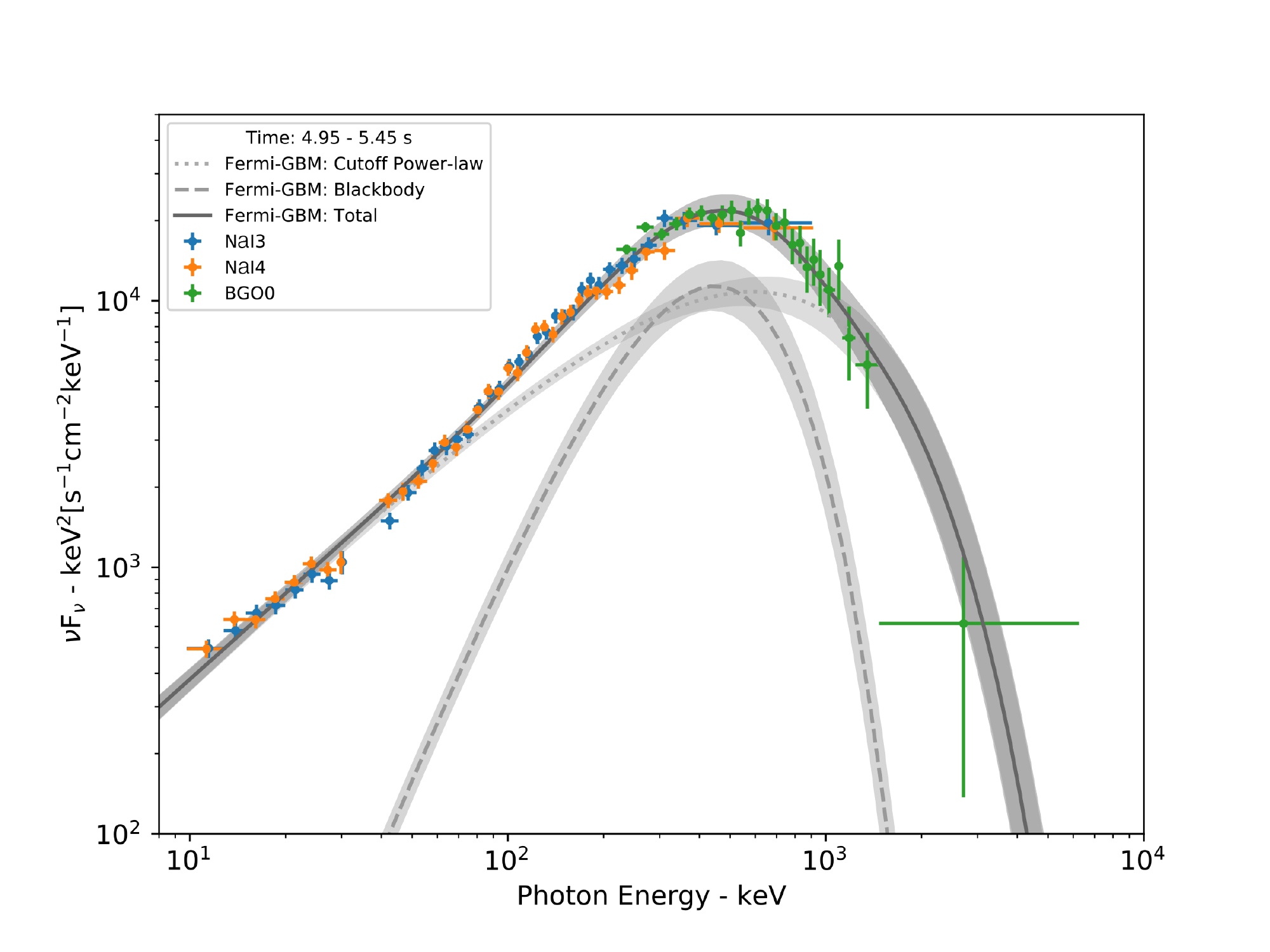}
\caption{Spectrum from $4.95$~s to $5.45$~s. The spectrum includes data from \emph{Fermi}-GBM (2 NaI and 1 BGO detector). The fitting is presented by a solid line, including the components of a Planck blackbody function indicated by a dashed line and a cutoff power law indicated by a dotted line.}
\label{fig:190114C_spectrum_gbm}
\end{figure*}

GRB spectra are known to evolve over different pulses, or even within a pulse. The time-integrated spectral analysis, therefore, must be replaced by the time-resolved spectral analysis to study the GRB radiation mechanism in great detail. We next performed a time-resolved spectral analysis for the {\it Fermi}-GBM observations. Thanks to its high fluence of (4.436$\pm$0.005)$\times$10$^{-4}$ erg cm$^{-2}$ as the fifth-highest fluence GRB ever observed by {\it Fermi}-GBM, we were able to divide its $T_{90}$ duration (116~s) into $48$ slices, with each time bin containing enough photons to conduct a high-significance spectral analysis. We used the typical GRB spectral model, the Band model \citep{Band1993}, to fit the time-resolved spectra in each slice (see Table \ref{tab:190114C}). We found that the low-energy photon index $\alpha$ exhibits a wide-spread temporal variability (-0.14 to -1.99), and the majority of $\alpha$ values in the first two pulses are harder than the typical value of $\alpha$ defined by the synchrotron line of death ($\alpha$=-2/3, \citealt{Preece1998}), suggesting a significant contribution from thermal emission from the fireball photosphere\citep{Ryde2010, Meszaros2000}. The majority of the high energy photon index $\beta$ values are not well-constrained, indicating that the CPL model is preferred in comparison with the Band model (Table \ref{tab:190114C}). The violation of the synchrotron limit encourages us to search for an additional thermal component. To search for the best model to characterise the spectral shape of the burst, we attempted to fit the time-resolved spectra in each slice with both the CPL and the CPL+BB models. The DIC of the CPL+BB model is at least 10 and can be hundreds less than the CPL model, indicating that adding a thermal component improves the spectral fitting greatly ($\rm \Delta DIC>10$, \citealt{Acuner2020}). The CPL+BB (CPL: Cutoff Power-law, BB: Blackbody) model\citep{Ryde2005, Battelino2007} gives a better fit in comparison with the CPL (see Table \ref{tab:CPLvsCPLBB}), Band and other models from $T_{0}$+$0.55$~s to $T_{0}$+$1.93$~s in $P_{1}$ (includes 8 slices, hereafter $P^{\rm th}_{1}$) and from $T_{0}$+$2.45$~s to $T_{0}$+$5.69$~s in $P_{2}$ (includes 16 slices, hereafter $P^{\rm th}_{2}$) based on the deviance information criterion (DIC). $P^{\rm th}_{1}$ and $P^{\rm th}_{2}$ correspond to the peak flux of the $P_{1}$ and $P_{2}$, respectively, which precisely correspond to the epochs when the power-law indices $\alpha$ of the single CPL fits are beyond the limits of the synchrotron line of death\citep{Preece1998}, i.e. $\alpha > -2/3$, indicating of the existence of a thermal component\citep{Ryde2004, Ghirlanda2007}. An example of an $\nu F_{\nu}$ spectrum for one time slice ($4.95$~s--$5.45$~s) with the CPL+BB model giving the best fit is displayed in Figure \ref{fig:190114C_spectrum_gbm}. Both $P_{1}$ and $P_{2}$ include non-thermal and sub-dominant thermal components. The thermal components observed in GRB 190114C exhibit pulse-wise temporal properties, i.e., those in $P_{1}$ and $P_{2}$ evolve independently over their pulse durations (Figure \ref{fig:temporal_KT_Gamma}). Such a feature provides a unique opportunity to study the photosphere properties at distinctly different epochs of central engine activities.

The time-resolved analysis shows that almost all the low-energy photon index $\alpha$ values of the CPL-only fits in $P_{\rm AO}$ are much softer than those in $P_{1}$ and $P_{2}$ (Figure \ref{fig:190114873_lc}), suggesting that the emission has a different origin. The index $\alpha$ gradually decreases toward $-2$, a typical value for synchrotron radiation, which indicates that the fireball has entered the afterglow phase. So we set the beginning of the epoch as the deceleration time when the mass of the ambient medium collected by the shockwave is comparable to $1/\Gamma$ of fireball energy\citep{Meszaros1993,Sari1999b}.

\subsection{Photosphere Properties}\label{sec:photosphere}

\begin{figure*}[!t]
\centering
\includegraphics[width=1\textwidth]{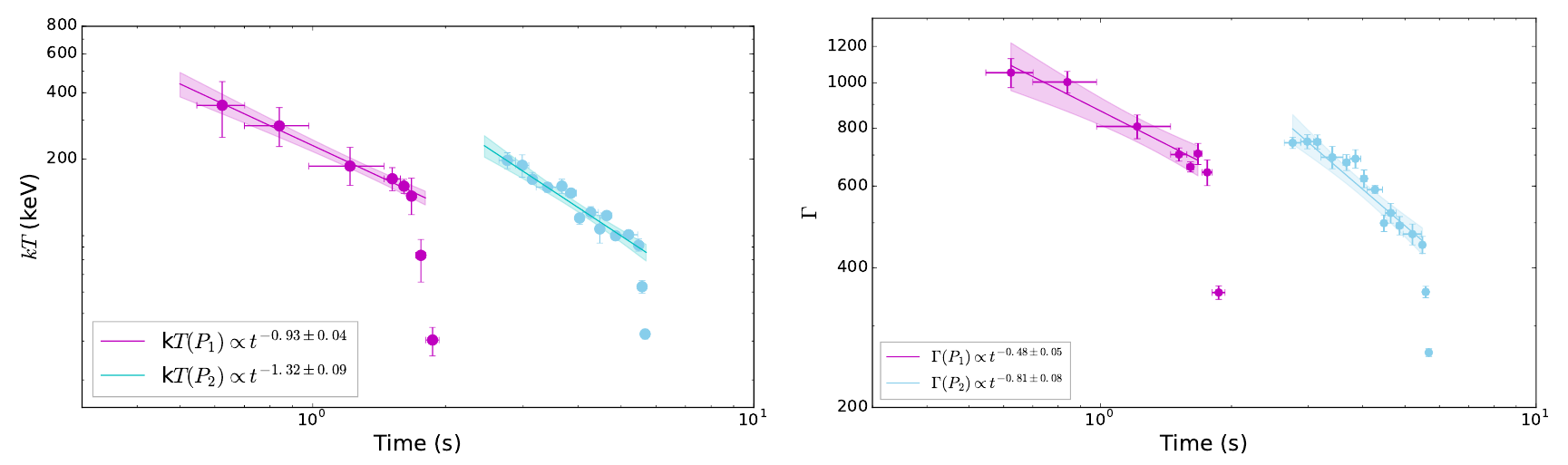}
\caption{Temporal evolution of the temperature k$T$ (left panel), and bulk Lorentz factor $\Gamma$ (right panel). The data points indicated by pink and cyan colours represent the two different pulses. Solid lines are the best power-law fits to the data for $P_{1}$ and $P_{2}$ excluding several points during the drop, and shaded areas are their 2-$\sigma$ (95\% confidence interval) regions. The derived time-resolved evolution of $\Gamma$ is based on the photosphere properties under the framework of the traditional method\citep{Peer2007}.} \label{fig:temporal_KT_Gamma}
\end{figure*}

We compared the properties of the thermal components identified in $P_{1}$ and $P_{2}$. The evolutions of the characteristic temperatures (k$T$) in $P_{1}$ and $P_{2}$ follow distinctly broken power-law decays: a smooth decay of the temperature followed by a fast drop (see the left panel in Figure \ref{fig:temporal_KT_Gamma}). The temporal feature in each pulse is consistent with the typical observations that showed a temperature evolution with a broken power law in time\citep{Ryde2004, Ryde2009}, but such a feature in two independent pulses within one burst has never been identified in previous observations. The temporal behaviours showing different decay indices between two different pulses within a single GRB suggest that the GRB central engine ejects distinct independent jet components during its active phase. We note that several GRBs with statistically significant thermal components have been observed by BATSE, Konus, {\it Swift}, and {\it Fermi} before\citep{Ryde2010,Guiriec2011,Guiriec2013,Axelsson2012}. However, they are either single-pulse bursts (e.g. GRB 110721A, \citealt{Axelsson2012}) or highly overlapping multi-pulse bursts (e.g. GRB 090902B, \citealt{Ryde2010}), or their thermal emission component is not strong enough (e.g. GRB 100724B, \citealt{Guiriec2011}), so that the photosphere properties could not be studied in detail among distinct pulses. The unique advantages of GRB 190114C, i.e. its low redshift, high fluence, several well-separated pulses in one single GRB, and strong thermal component, make such a study possible.

\begin{figure*}[!t]
\includegraphics[scale=0.8]{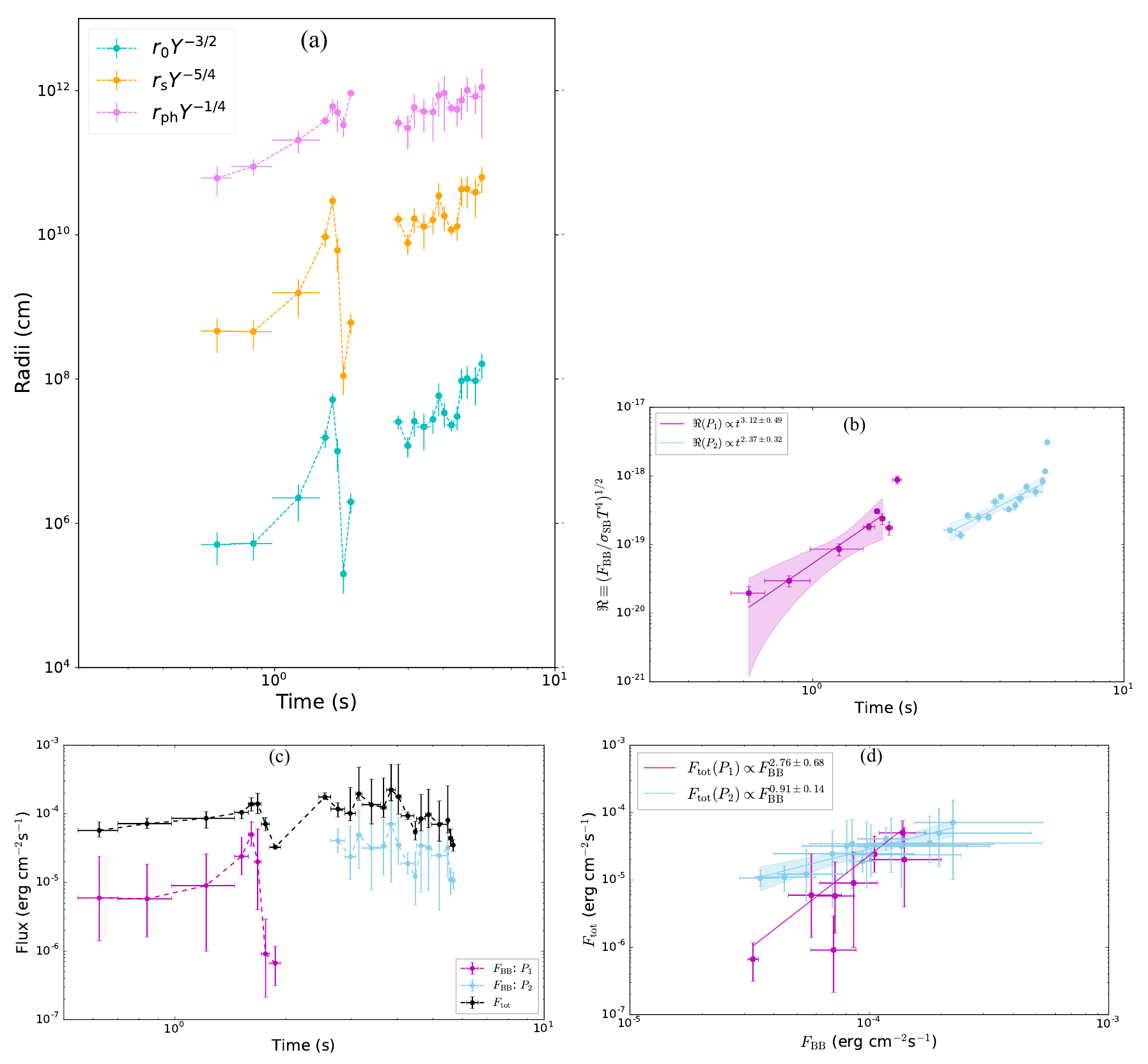}
\caption{Pulse-wise properties of GRB 190114C. {\bf(a)} Temporal evolution of the photospheric radius $r_{\rm ph}$ (violet), saturation radius $r_{\rm s}$ (orange), and nozzle radius $r_{\rm 0}$ (cyan). {\bf(b)} Temporal evolution of the parameter $\Re$. {\bf(c)} Temporal evolution of the BB energy flux ($F_{\rm BB}$) and total energy flux ($F_{\rm tot}$). {\bf(d)} The total energy flux ($F_{\rm tot}$) versus the BB energy flux ($F_{\rm BB}$). Same color notation as in Figure \ref{fig:temporal_KT_Gamma}.}
\label{fig:pulse_temporal}
\end{figure*}

Within the framework of the standard fireball photosphere model\citep{Peer2007}, we can infer the photosphere characteristics and the ratio of thermal to non-thermal emission to obtain information on the jet properties, such as the bulk Lorentz factor $\Gamma$ and the initial size of the jet $r_{0}$. Figure \ref{fig:temporal_KT_Gamma} and Figure \ref{fig:pulse_temporal} show the evolution of the bulk Lorentz factor $\Gamma$ and the parameter $\Re$, respectively; where $\Re$ is the effective transverse size of the emitting region\citep{Ryde2009}. They exhibit similar temporal behaviors in $P_{1}$ and $P_{2}$, i.e., a broken power-law evolution behavior, with $\Re$ increasing with time and $\Gamma$ decreasing over time. The comparison of the properties of a global view as well as the best-fitting results of the relevant parameters is summarized in Table \ref{tab:Pulse}.

The derived Lorentz factors and the photosphere radii exhibit systematic variations, with the Lorentz factor decreasing from $\sim 1000$ to $\sim 200$ (Figure \ref{fig:temporal_KT_Gamma}), and the photosphere radius varying on the order of $10^{12}$~cm (Figure \ref{fig:pulse_temporal}a). This is likely related to the behaviour of the GRB central engine. The decay of $\Gamma$ in $P_1$ and $P_2$ is consistent with the expectation that faster ejecta from the engine tends to reach the photosphere earlier than slower ejecta, and the rapid decline at the end of each episode may be related to the abrupt cessation of the engine activity\citep{Li2021a}, with the decay slope defined by the ebbing ejection rate of the central engine. Since the Lorentz factor range is not very wide, it is expected that the deceleration of the fireball is essentially prompt without a significant energy injection phase due to the pile-up of the slow materials. This is consistent with the power-law decay with time of the multi-wavelength afterglow emission from the source\citep{MAGICCollaboration2019,Wang2019,Wang2019b}.

\subsection{Application to GRB 190114C with Our New Method}

In short, GRB 190114C is unique in terms of the following aspects. (1) It has three well-separated emission episodes, which can be defined as the first, second, and third pulses. (2) The emission of the first two main pulses consists of two strong thermally-subdominated episodes, which independently exhibit similar temporal properties. (3) The first two pulses (thermal) and the third pulse (non-thermal) have distinct spectral properties. (4) The thermal component has a thermal to total flux ratio of 21$^{+6}_{-4}$\%, which is the second-highest among the GRBs observed with {\it Fermi}-GBM so far (the highest one is observed in GRB 090902B, with thermal flux ratio $\sim$ 70\%). (5) Strong TeV emission was observed, setting the record of one of the highest photon energy in any GRB\citep{MAGICCollaboration2019}. The two well-separated pulses with independent and analogous thermal component evolution patterns make this extraordinarily bright GRB a unique event to study the jet composition and evolution of the photospheric properties in a single GRB. We note that several interesting cases have been observed in the past. For example, in some GRBs, a hot fireball jet characterized by a quasi-thermal Planck-like spectrum was observed (e.g. GRB 090902B, \citealt{Abdo2009}). In many other GRBs, a Poynting-flux-dominated outflow characterized by a Band (or cutoff power-law)-only function may also be observed (e.g., GRB 080916C,\citealt{Abdo2009a} and GRB 130427A, \citealt{Preece2016}). More interestingly, we may also observe a hybrid jet characterized by either a two-component spectral scenario (composed of a non-thermal component and a thermal component simultaneously, e.g., GRB 110721A, \citealt{Axelsson2012,Gao2015}, or a transition from fireball to Poynting-flux-dominated outflow within a single GRB (e.g., GRB 160625B, \citealt{Ryde2011,Zhang2018a,Li2019a}). However, GRB 190114C presented unique information not available before.

The above-mentioned two methods (see Section \ref{sec:photosphere}) of measuring Lorentz factors both rely on some unknown parameters. By combining the photosphere data in $P_{1}$ and $P_{2}$ and the afterglow data in $P_{\rm AO}$, one can discriminate various energy components in the fireball in an essentially parameter-independent way\citep{Zhang2021}. A systematic search in previously detected GRBs did not reveal a single case showing both a significant photosphere signature and an afterglow deceleration signature\citep{Zhang2021}. GRB 190114C, therefore, provides the first case in which the determinateness of fireball parameters\citep{Zhang2021} can be carried out. We perform a time-integrated spectral fit to the prompt emission spectra of $P_{1}$ and $P_{2}$ (0.55 - 1.93 s and 2.45-5.69 s) with the CPL+BB model and derive the observed properties (including both the thermal and non-thermal components) of the fireball as shown in Table \ref{tab:global}. Following \cite{Zhang2021} (for details see Section \ref{sec:newmodel}), we can derive the following physical parameters of a GRB fireball (Table \ref{tab:global}): initial dimensionless specific enthalpy $\eta =854 \pm 38$, bulk Lorentz factor at the photosphere $\Gamma_{\rm ph} = 833 \pm 38$, bulk Lorentz factor before deceleration $\Gamma_0 = 719 \pm 59$, and fireball isotropic-equivalent mass loading $M_{\rm iso}=(1.7\pm 0.4) \times 10^{-3} M_\odot$. This gives a direct measurement of the fireball radiative efficiency $\eta_\gamma = (15.8 \pm 5.4) \%$. This measured efficiency has much smaller uncertainties than the values derived for previous GRBs using afterglow modeling\citep{Zhang2007b}. A high fireball radiative efficiency has been theorized in the past\citep{Meszaros2000,Kobayashi2001}. Our measured $\eta_\gamma \sim 16\%$ suggests that a GRB fireball can indeed emit both thermal and non-thermal gamma-rays efficiently. We also find that the derived bulk Lorentz factors measured during the prompt emission phase ($\Gamma=854\pm38$) are slightly higher than the bulk Lorentz factor measured at the deceleration radius ($\Gamma_{0}=719\pm59$). This is fully consistent with the picture described by the GRB fireball model in which a fraction of the kinetic energy is dissipated during the prompt emission phase.

\begin{figure}
\centering
\includegraphics[scale=0.6]{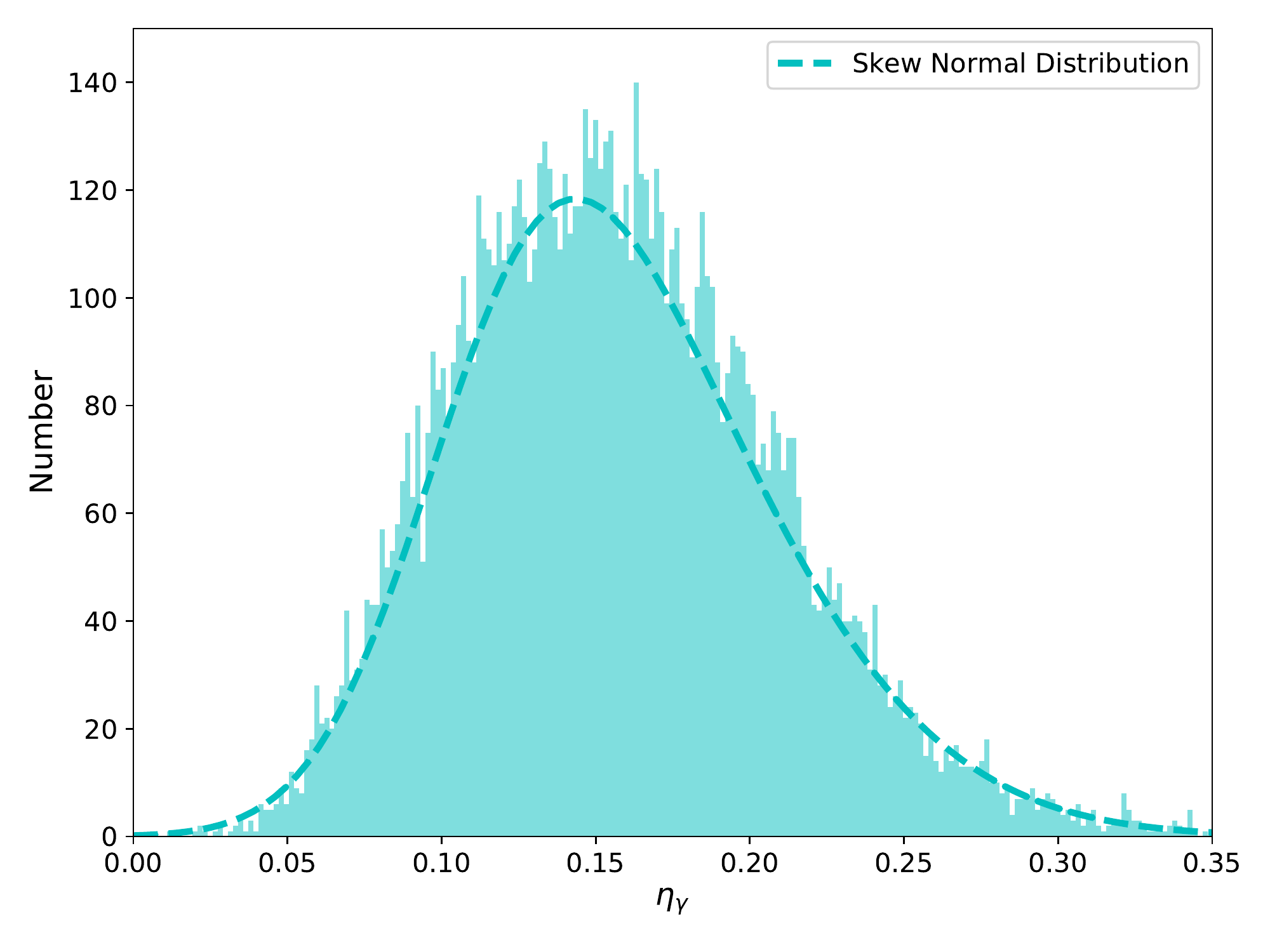}
\caption{Histogram of $\eta_\gamma$ from 10000 Monte Carlo samplings, each bin corresponds to the $\eta_\gamma$ interval of $0.003$. The histogram is fitted by a skew-normal distribution function.}
\label{fig:skewnorm}
\end{figure}

To solve the above equations described in Section \ref{sec:newmodel}, we apply the Monte-Carlo method to obtain the mean and the uncertainty for the measured values and the uncertainties of $E_{\rm th,iso}, E_{\rm nth,iso}, F_{\rm BB}^{\rm obs}, F_{\rm \gamma}^{\rm obs}, kT^{\rm obs}$. We sample, for each of them, $10000$ times following the normal distribution. We set a range of $6-10$~s for $t_{\rm dec}$ while of $0.5-1.5$~cm$^{-3}$ for $n$. We obtain the values of $\eta$, $\Gamma_{\rm ph}, \Gamma_0, M_{\rm iso}$, $E_{\rm k,iso}$, $E_{\rm tot,iso}$ and $\eta_\gamma$, computed from the $10000$ samples by the above equations and find they can be fitted by skew-normal distributions, see e.g. in Figure \ref{fig:skewnorm}, from which the mean and the asymmetrical uncertainties are derived. The average values based on the two thermal episodes of $P^{\rm th}_{1}$ (from 0.55 to 1.93 s) and $P^{\rm th}_{2}$  (from 2.45 to 5.69 s) are given in Table \ref{tab:global}. All the measured quantities are presented in the upper panel, and all the derived parameters are presented in the lower panel.

\subsection{Further Estimate of the Energy Fractions Assigned to Electrons ($\epsilon_{e}$) and Magnetic ($\epsilon_{B}$) fields}

Once $E_{\rm k}$ is precisely obtained from the observational data using our new methods discussed above, one can estimate the energy fractions assigned to electrons ($\epsilon_{e}$) and magnetic fields ($\epsilon_{B}$) using afterglow models\citep{Zhang2007b}.

The isotropic blastwave kinetic energy ($E_{\rm K,iso}$) can also be measured from the afterglow emission (normal decay) using the {\it Swift}-XRT data. For a constant density interstellar medium (ISM), e.g.,~\cite{Schulze2011}, the characteristic synchrotron frequency and the cooling frequency of minimum-energy injected electrons, and therefore the peak spectral flux, can be given by~\cite{Sari1998,Yost2003,Zhang2007b}
\begin{equation}
\nu_{\rm m}=3.3 \times 10^{12} {\rm Hz} \left(\frac{p-2}{p-1} \right)^{2}(1+z)^{1/2} \varepsilon_{\rm B,-2}^{1/2}  \varepsilon_{\rm e,-1}^{2}E^{1/2}_{\rm K,iso,52}t^{-3/2}_{\rm d},
\end{equation}
\begin{equation}
\nu_{\rm c}=6.3 \times 10^{15} {\rm Hz} (1+z)^{-1/2} (1+Y)^{-2}\varepsilon_{\rm B,-2}^{-3/2}E^{-1/2}_{\rm K,iso,52}n^{-1} t^{-1/2}_{\rm d},
\label{eq:nu_c}
\end{equation}
\begin{equation}
F_{\nu, \rm max}=1.6 {\rm mJy} (1+z)D_{28}^{-2}\varepsilon_{\rm B,-2}^{1/2}E_{\rm K,iso,52}n^{-1},
\label{eq:max_Flux}
\end{equation}
where $p$ is the electron spectral distribution index, $\epsilon_{e}$ and $\epsilon_{B}$ are the energy fractions assigned to electrons and magnetic fields, $t_{\rm d}$ is the time in the observer frame in units of days, $D_{28}=D/10^{28}$, is the luminosity distance in units\footnote{The convention $Q=10^{x}Q_{x}$ is adopted in cgs units for all parameters throughout the paper.} of $10^{28}$ cm, $n$ is the number density in the constant density ambient medium, and
\begin{equation}
Y=\left[-1+(1+4\eta_{1}\eta_{2}\varepsilon_{\rm e}/\varepsilon_{\rm B})^{1/2}\right]/2,
\label{eq:Y}
\end{equation}
is the Inverse Compton (IC) parameter, where $\eta_{1}=\rm {min}[1,(\nu_{c}/\nu_{m})^{(2-p)/2}]$, $\eta_{2}=\rm {min}[1,(\nu_{\rm KN}/\nu_{c})^{(3-p)/2}]$ (for the slow cooling $\nu_{m}<\nu_{x}<\nu_{c}$ case) is a correction factor introduced by the Klein-Nishina effect, where $\nu_{\rm KN}$ is the Klein-Nishina frequency
\begin{equation}
\begin{split}
\nu_{\rm KN}=h^{-1}\Gamma m_{e} c^{2}\gamma^{-1}_{e,X}(1+z)^{-1}\simeq 2.4 \times 10^{15} {\rm Hz} (1+z)^{-3/4} E^{1/4}_{\rm K,iso,52} \varepsilon_{\rm B,-2}^{1/4} t^{-3/4}_{\rm d} \nu^{-1/2}_{18},
\end{split}
\end{equation}
where $h$ is Plancks constant and $\gamma_{e,X}$ is the electron Lorentz factor corresponding to the X-ray band emission.

The spectral regime can be determined by using the closure relation in the afterglow emission via the observed temporal ($\hat{\alpha}$) and spectral ($\hat{\beta}$) indices. The temporal index $\hat{\alpha}_{\rm XRT}$ is measured from the {\it Swift}-XRT lightcurve (see Figure \ref{fig:multi_lc_spectrum}a, and the corresponding spectral index $\hat{\beta}_{\rm XRT}= -(\Gamma_{\rm XRT}-1)=-0.93\pm 0.10$ ($\Gamma_{\rm XRT}$ is the photon spectral index) is available from the {\it Swift} online server~\citep{Evans2007, Evans2009}. Using the temporal and spectral indices, one can therefore determine that the X-ray emission in GRB 190114C is in the $\nu_{\rm m}<\nu_{\rm x}<\nu_{\rm c}$ regime. With the spectral regime known, the electron index $p$ can be derived using the observed temporal index: $p=(3-4\hat{\alpha}_{\rm XRT})/3=2.85 \pm 0.01$. 

In the case of $p>2$, and in the $\nu_{\rm m}<\nu_{\rm x}<\nu_{\rm c}$ regime, one can derive the X-ray band energy flux as 
\begin{equation}
\begin{split}
\nu F_{\nu}(\nu=10^{18} {\rm Hz})=F_{\nu, {\rm max}}
(\nu_{\rm m}/\nu_{\rm x})^{(p-1)/2}\\
=6.5\times10^{-13} {\rm erg s^{-1}cm^{-2}}D_{28}^{-2}(1+z)^{(p+3)/4}\\
\times f_{p} \varepsilon_{B,-2}^{(p+1)/4} \varepsilon_{e,-1}^{p-1} E^{(p+3)/4}_{\rm K,iso,52}n^{1/2}t_{d}^{(3-3p)/4}\nu_{18}^{(3-p)/2}.
\end{split}
\label{eq:E_k}
\end{equation}
This gives,
\begin{equation}
\begin{split}
E_{\rm K,iso,52}=\left[\frac{\nu F_{\nu}(\nu=10^{18} {\rm Hz})}{6.5\times10^{-13} \rm erg s^{-1}cm^{-2}}\right]^{4/(p+3)}\\
\times D_{28}^{8/(p+3)}(1+z)^{-1}t_{d}^{3(p-1)/(p+3)}\\
\times f_{p}^{-4/(p+3)}\varepsilon_{B,-2}^{-(p+1)/(p+3)}\varepsilon_{e,-1}^{4(1-p)/(p+3)}\\
\times n^{-2/(p+3)}\nu_{18}^{2(p-3)/(p+3)},
\end{split}
\label{eq:E_k}
\end{equation}
where $\nu F_{\nu}$ ($\nu=10^{18})$ Hz is the energy flux at frequency $10^{18}$ Hz in units of ${\rm erg \ s^{-1} \ cm^{-2}}$, and 
\begin{equation}
f_{p}=6.73 \left(\frac{p-2}{p-1}\right)^{p-1}\left(3.3\times10^{-6}\right)^{(p-2.3)/2}.
\end{equation} 
is a function of the electron power-law index $p$.

Simultaneously solving Eq. \ref{eq:Y} and Eq. \ref{eq:E_k}, with the IC parameter $Y^{\rm IC}(=E_{\rm GeV}/E_{\rm MeV}$) constrained from the observations in GRB 190114C, e.g. $Y^{\rm IC}=0.75$~\citep[e.g.,][]{Wang2019b}, and the using episodes of $P^{\rm th}_{1}$ and $P^{\rm th}_{2}$, we obtain $\epsilon_{\rm B}$ and $\epsilon_{\rm e}$ during the time interval of $P_{\rm AO}$,
\begin{eqnarray}
\label{n} \left\{\begin{array}{ll}
\epsilon_{\rm e,-1}=1.11\pm0.01,\\
\epsilon_{\rm B,-2}=0.05\pm0.01,\\
\end{array} \right.
\end{eqnarray}

Knowing values of $\epsilon_{\rm B}$ and $\epsilon_{\rm e}$, we can also solve for $\nu_{\rm m}$, $\nu_{\rm c}$, and $\nu_{\rm KN}$, 
\begin{eqnarray}
\label{n} \left\{\begin{array}{ll}
\nu_{\rm m}=(1.30\pm0.82)\times 10^{17} \rm Hz,\\
\nu_{\rm c}=(4.44\pm0.66)\times 10^{17} \rm Hz,\\
\nu_{\rm KN}=(6.55\pm0.16)\times 10^{17} \rm Hz\\
\end{array} \right.
\end{eqnarray}

\section{Conclusions} \label{sec:conclusion}

In this paper, using the photosphere data observed in $P_{1}$ and $P_{2}$, and the early afterglow data observed in $P_{\rm AO}$, as well as a measured redshift, we apply the method proposed in \cite{Zhang2021} to directly dissect the GRB fireball energy budget and therefore to measure GRB radiative efficiency for GRB 190114C.  

We first performed a detailed time-integrated and time-resolved spectral analysis for the \emph{Fermi}-GBM observations by using various GRB spectral models. Its prompt emission consists of three well-separated pulses. We found a strong thermal component observed in the first two emission pulses. The spectra in $P_1$ and $P_2$ are best fitted by a two-component scenario, with a non-thermal Band-like component accompanied by a thermal blackbody component. The thermal component has a thermal to total flux ratio of 21$^{+6}_{-4}$\%. Such a strong thermal component found in the time-resolved spectral analysis between well-separated pulses in GRB 190114C gives a good opportunity to study the photospheric properties, and allows us for the first time to study a fine time-resolved spectral analysis and track the blackbody evolution among the different pulses in a single GRB. Indeed, we found two well-separated thermal pulses evolving independently and analogically, inferred from their observational and physical parameters derived from the fireball model. Such independent and analogical pulse-wise thermal properties in GRB 190114C are the first case found in GRB history, which strongly supports the evidence of a shell-like structure during the prompt emission phase. We also found that starting from the third pulse ($P_{3}$) and extending to the entire afterglow, the spectra are all non-thermal, the synchrotron plus Compton up-scattering model well interprets the observation , and consequently the fireball parameters are obtained. More interestingly, the onset signature of afterglow emission corresponding to the deceleration of the fireball was observed to be from $T_{0}$+$6$~s to $T_{0}$+$15$~s in $P_{\rm AO}$ due to the fact that multiple pieces of observational evidence (e.g., flux, energy band, and power-law index) are consistent with external shock emissions. By incorporating the thermal ($P_{1}$ and $P_{2}$) and the non-thermal ($P_{\rm AO}$) observations, as well as the photosphere and the synchrotron radiative mechanisms, we directly derived the fireball energy budget with little dependence on hypothetical parameters \citep{Zhang2021} and to measure a $\sim$16\% radiative efficiency for this GRB.

With the fireball parameters that have been determined, the isotropic kinetic energy\citep{Panaitescu2001} of the fireball at the afterglow phase is measured as $E_{\rm k,iso} = (1.6 \pm 0.7) \times 10^{54}$ erg. This allows us to make use of this prompt-emission-measured $E_{\rm k,iso}$ in the afterglow model to constrain shock microphysics parameters. Using broad-band afterglow data, we can derive an electron injection power law index $p \simeq 2.85$ and the inverse Compton parameter $Y^{\rm IC} \sim 0.75$. This leads to the determination of the two equipartition parameters of electrons and magnetic fields: $\epsilon_e = (1.11 \pm 0.01) \times 10^{-1}$ and $\epsilon_{\rm B} =(0.5 \pm 0.1) \times 10^{-3}$. These parameters are usually poorly constrained in other GRBs unless there is complete multi-wavelength afterglow data\citep{Panaitescu2002}. We are able to measure these values more precisely, and they are also broadly consistent with the afterglow modeling of the event\citep{MAGICCollaboration2019}.

\acknowledgments

We thank the anonymous referee for his/her valuable comments and suggestions. We also thank Damien B\'egu\'e, H\"usne Dereli-B\'egu\'e, Michael S. Briggs, Xue-Feng Wu, Zi-Gao Dai, Ye-Fei Yuan, Yi-Fu Cai, En-Wei Liang, Remo Ruffini, and ICRANet members for many helpful discussions on GRB physics and phenomena. In particular, LL would like to dedicate this piece to the memory of Dr. Magnus Axelsson, a close colleague who passed away recently and was one of its main contributors. AJC-T acknowledges financial support from the State Agency for Research of the Spanish MCIU through the ``Center of Excellence Severo Ochoa" award to the Instituto de Astrofísica de Andalucía (SEV-2017-0709). DAK acknowledges support from Spanish National Research Project RTI2018-098104-J-I00 (GRBPhot). We also acknowledge the use of public data from the {\it Fermi} Science Support Center (FSSC) and the UK {\it Swift} Science Data Center.
 
\bibliography{Myreferences.bib}	

\begin{thebibliography}{}
\expandafter\ifx\csname natexlab\endcsname\relax\def\natexlab#1{#1}\fi

\bibitem[{{Abdo} {et~al.}(2009){Abdo}, {Ackermann}, {Ajello}, {Asano},
  {Atwood}, {Axelsson}, {Baldini}, {Ballet}, {Barbiellini}, {Baring},
  {Bastieri}, {Bechtol}, {Bellazzini}, {Berenji}, {Bhat}, {Bissaldi},
  {Blandford}, {Bloom}, {Bonamente}, {Borgland}, {Bouvier}, {Bregeon}, {Brez},
  {Briggs}, {Brigida}, {Bruel}, {Burgess}, {Burrows}, {Buson}, {Caliandro},
  {Cameron}, {Caraveo}, {Casandjian}, {Cecchi}, {{\c C}elik}, {Chekhtman},
  {Cheung}, {Chiang}, {Ciprini}, {Claus}, {Cohen-Tanugi}, {Cominsky},
  {Connaughton}, {Conrad}, {Cutini}, {d'Elia}, {Dermer}, {de Angelis}, {de
  Palma}, {Digel}, {Dingus}, {Silva}, {Drell}, {Dubois}, {Dumora}, {Farnier},
  {Favuzzi}, {Fegan}, {Finke}, {Fishman}, {Focke}, {Fortin}, {Frailis},
  {Fukazawa}, {Funk}, {Fusco}, {Gargano}, {Gehrels}, {Germani}, {Giavitto},
  {Giebels}, {Giglietto}, {Giordano}, {Glanzman}, {Godfrey}, {Goldstein},
  {Granot}, {Greiner}, {Grenier}, {Grove}, {Guillemot}, {Guiriec}, {Hanabata},
  {Harding}, {Hayashida}, {Hays}, {Horan}, {Hughes}, {Jackson},
  {J{\'o}hannesson}, {Johnson}, {Johnson}, {Johnson}, {Kamae}, {Katagiri},
  {Kataoka}, {Kawai}, {Kerr}, {Kippen}, {Kn{\"o}dlseder}, {Kocevski}, {Komin},
  {Kouveliotou}, {Kuss}, {Lande}, {Latronico}, {Lemoine-Goumard}, {Longo},
  {Loparco}, {Lott}, {Lovellette}, {Lubrano}, {Madejski}, {Makeev},
  {Mazziotta}, {McBreen}, {McEnery}, {McGlynn}, {Meegan}, {M{\'e}sz{\'a}ros},
  {Meurer}, {Michelson}, {Mitthumsiri}, {Mizuno}, {Moiseev}, {Monte},
  {Monzani}, {Moretti}, {Morselli}, {Moskalenko}, {Murgia}, {Nakamori},
  {Nolan}, {Norris}, {Nuss}, {Ohno}, {Ohsugi}, {Omodei}, {Orlando}, {Ormes},
  {Paciesas}, {Paneque}, {Panetta}, {Pelassa}, {Pepe}, {Pesce-Rollins},
  {Petrosian}, {Piron}, {Porter}, {Preece}, {Rain{\`o}}, {Rando}, {Rau},
  {Razzano}, {Razzaque}, {Reimer}, {Reimer}, {Reposeur}, {Ritz}, {Rochester},
  {Rodriguez}, {Roming}, {Roth}, {Ryde}, {Sadrozinski}, {Sanchez}, {Sander},
  {Saz Parkinson}, {Scargle}, {Schalk}, {Sgr{\`o}}, {Siskind}, {Smith},
  {Spinelli}, {Stamatikos}, {Stecker}, {Stratta}, {Strickman}, {Suson},
  {Swenson}, {Tajima}, {Takahashi}, {Tanaka}, {Thayer}, {Thayer}, {Thompson},
  {Tibaldo}, {Torres}, {Tosti}, {Tramacere}, {Uchiyama}, {Uehara}, {Usher},
  {van der Horst}, {Vasileiou}, {Vilchez}, {Vitale}, {von Kienlin}, {Waite},
  {Wang}, {Wilson-Hodge}, {Winer}, {Wood}, {Yamazaki}, {Ylinen}, \&
  {Ziegler}}]{Abdo2009}
{Abdo}, A.~A., {Ackermann}, M., {Ajello}, M., {et~al.} 2009, {The Astrophysical
  Journal Letters}, 706, L138

\bibitem[{Abdo {et~al.}(2009)Abdo, Ackermann, Arimoto, Asano, Atwood, Axelsson,
  Baldini, Ballet, Band, Barbiellini, Baring, Bastieri, Battelino, Baughman,
  Bechtol, Bellardi, Bellazzini, Berenji, Bhat, Bissaldi, Blandford, Bloom,
  Bogaert, Bogart, Bonamente, Bonnell, Borgland, Bouvier, Bregeon, Brez,
  Briggs, Brigida, Bruel, Burnett, Burrows, Busetto, Caliandro, Cameron,
  Caraveo, Casandjian, Ceccanti, Cecchi, Celotti, Charles, Chekhtman, Cheung,
  Chiang, Ciprini, Claus, Cohen-Tanugi, Cominsky, Connaughton, Conrad,
  Costamante, Cutini, DeKlotz, Dermer, de~Angelis, de~Palma, Digel, Dingus,
  do~Couto~e Silva, Drell, Dubois, Dumora, Edmonds, Evans, Fabiani, Farnier,
  Favuzzi, Finke, Fishman, Focke, Frailis, Fukazawa, Funk, Fusco, Gargano,
  Gasparrini, Gehrels, Germani, Giebels, Giglietto, Giommi, Giordano, Glanzman,
  Godfrey, Goldstein, Granot, Greiner, Grenier, Grondin, Grove, Guillemot,
  Guiriec, Haller, Hanabata, Harding, Hayashida, Hays, Morata, Hoover, Hughes,
  Johannesson, Johnson, Johnson, Johnson, Johnson, Kamae, Katagiri, Kataoka,
  Kavelaars, Kawai, Kelly, Kennea, Kerr, Kippen, Knodlseder, Kocevski, Kocian,
  Komin, Kouveliotou, Kuehn, Kuss, Lande, Landriu, Larsson, Latronico,
  Lavalley, Lee, Lee, Lemoine-Goumard, Lichti, Longo, Loparco, Lott,
  Lovellette, Lubrano, Madejski, Makeev, Marangelli, Mazziotta, McBreen,
  McEnery, McGlynn, Meegan, M{\'e}sz{\'a}ros, Meurer, Michelson, Minuti,
  Mirizzi, Mitthumsiri, Mizuno, Moiseev, Monte, Monzani, Moretti, Morselli,
  Moskalenko, Murgia, Nakamori, Nelson, Nolan, Norris, Nuss, Ohno, Ohsugi,
  Okumura, Omodei, Orlando, Ormes, Ozaki, Paciesas, Paneque, Panetta, Parent,
  Pelassa, Pepe, Perri, Pesce-Rollins, Petrosian, Pinchera, Piron, Porter,
  Preece, Raino, Ramirez-Ruiz, Rando, Rapposelli, Razzano, Razzaque, Rea,
  Reimer, Reimer, Reposeur, Reyes, Ritz, Rochester, Rodriguez, Roth, Ryde,
  Sadrozinski, Sanchez, Sander, Parkinson, Scargle, Schalk, Segal, Sgro,
  Shimokawabe, Siskind, Smith, Smith, Spandre, Spinelli, Stamatikos, Starck,
  Stecker, Steinle, Stephens, Strickman, Suson, Tagliaferri, Tajima, Takahashi,
  Takahashi, Tanaka, Tenze, Thayer, Thayer, Thompson, Tibaldo, Torres, Tosti,
  Tramacere, Turri, Tuvi, Usher, Van~der Horst, Vigiani, Vilchez, Vitale, von
  Kienlin, Waite, Williams, Wilson-Hodge, Winer, Wood, Wu, Yamazaki, Ylinen,
  Ziegler, Collaboration, \& Collaboration}]{Abdo2009a}
Abdo, A.~A., Ackermann, M., Arimoto, M., {et~al.} 2009, Science, 323, 1688

\bibitem[{{Acuner} {et~al.}(2020){Acuner}, {Ryde}, {Pe'er}, {Mortlock}, \&
  {Ahlgren}}]{Acuner2020}
{Acuner}, Z., {Ryde}, F., {Pe'er}, A., {Mortlock}, D., \& {Ahlgren}, B. 2020,
  {The Astrophysical Journal}, 893, 128

\bibitem[{{Ajello} {et~al.}(2020){Ajello}, {Arimoto}, {Axelsson}, {Baldini},
  {Barbiellini}, {Bastieri}, {Bellazzini}, {Berretta}, {Bissaldi}, {Blandford},
  {Bonino}, \& {De Pasquale}}]{ajello2020}
{Ajello}, M., {Arimoto}, M., {Axelsson}, M., {et~al.} 2020, {The Astrophysical
  Journal}, 890, 9

\bibitem[{{Alexander} {et~al.}(2019){Alexander}, {Laskar}, {Berger}, {Mundell},
  \& {Margutti}}]{Alexander2019}
{Alexander}, K.~D., {Laskar}, T., {Berger}, E., {Mundell}, C.~G., \&
  {Margutti}, R. 2019, GRB Coordinates Network, 23726, 1

\bibitem[{{Axelsson} {et~al.}(2012){Axelsson}, {Baldini}, {Barbiellini},
  {Baring}, {Bellazzini}, {Bregeon}, {Brigida}, {Bruel}, {Buehler},
  {Caliandro}, {Cameron}, {Caraveo}, {Cecchi}, {Chaves}, {Chekhtman}, {Chiang},
  {Claus}, {Conrad}, {Cutini}, {D'Ammando}, {de Palma}, {Dermer}, {Silva},
  {Drell}, {Favuzzi}, {Fegan}, {Ferrara}, {Focke}, {Fukazawa}, {Fusco},
  {Gargano}, {Gasparrini}, {Gehrels}, {Germani}, {Giglietto}, {Giroletti},
  {Godfrey}, {Guiriec}, {Hadasch}, {Hanabata}, {Hayashida}, {Hou}, {Iyyani},
  {Jackson}, {Kocevski}, {Kuss}, {Larsson}, {Larsson}, {Longo}, {Loparco},
  {Lundman}, {Mazziotta}, {McEnery}, {Mizuno}, {Monzani}, {Moretti},
  {Morselli}, {Murgia}, {Nuss}, {Nymark}, {Ohno}, {Omodei}, {Pesce-Rollins},
  {Piron}, {Pivato}, {Racusin}, {Rain{\`o}}, {Razzano}, {Razzaque}, {Reimer},
  {Roth}, {Ryde}, {Sanchez}, {Sgr{\`o}}, {Siskind}, {Spandre}, {Spinelli},
  {Stamatikos}, {Tibaldo}, {Tinivella}, {Usher}, {Vandenbroucke}, {Vasileiou},
  {Vianello}, {Vitale}, {Waite}, {Winer}, {Wood}, {Burgess}, {Bhat},
  {Bissaldi}, {Briggs}, {Connaughton}, {Fishman}, {Fitzpatrick}, {Foley},
  {Gruber}, {Kippen}, {Kouveliotou}, {Jenke}, {McBreen}, {McGlynn}, {Meegan},
  {Paciesas}, {Pelassa}, {Preece}, {Tierney}, {von Kienlin}, {Wilson-Hodge},
  {Xiong}, \& {Pe'er}}]{Axelsson2012}
{Axelsson}, M., {Baldini}, L., {Barbiellini}, G., {et~al.} 2012, {The
  Astrophysical Journal Letters}, 757, L31

\bibitem[{{Band} {et~al.}(1993){Band}, {Matteson}, {Ford}, {Schaefer},
  {Palmer}, {Teegarden}, {Cline}, {Briggs}, {Paciesas}, {Pendleton}, {Fishman},
  {Kouveliotou}, {Meegan}, {Wilson}, \& {Lestrade}}]{Band1993}
{Band}, D., {Matteson}, J., {Ford}, L., {et~al.} 1993, {The Astrophysical
  Journal}, 413, 281

\bibitem[{{Battelino} {et~al.}(2007){Battelino}, {Ryde}, {Omodei}, \&
  {Band}}]{Battelino2007}
{Battelino}, M., {Ryde}, F., {Omodei}, N., \& {Band}, D.~L. 2007, in American
  Institute of Physics Conference Series, Vol. 921, The First GLAST Symposium,
  ed. S.~{Ritz}, P.~{Michelson}, \& C.~A. {Meegan}, 478--479

\bibitem[{{B{\'e}gu{\'e}} \& {Iyyani}(2014)}]{Begue2014}
{B{\'e}gu{\'e}}, D., \& {Iyyani}, S. 2014, {The Astrophysical Journal}, 792, 42

\bibitem[{{Bikmaev} {et~al.}(2019){Bikmaev}, {Irtuganov}, {Sakhibullin},
  {Burenin}, {Pavlinsky}, {Sunyaev}, {Khamitov}, {Ozdemir}, \&
  {Gogus}}]{Bikmaev2019}
{Bikmaev}, I., {Irtuganov}, E., {Sakhibullin}, N., {et~al.} 2019, GRB
  Coordinates Network, 23766, 1

\bibitem[{{Bolmer} \& {Schady}(2019)}]{Bolmer2019}
{Bolmer}, J., \& {Schady}, P. 2019, GRB Coordinates Network, 23702, 1

\bibitem[{Cash(1979)}]{Cash1979}
Cash, W. 1979, ApJ, 228, 939

\bibitem[{{Castro-Tirado} {et~al.}(2019){Castro-Tirado}, {Hu},
  {Fernandez-Garcia}, {Valeev}, {Sokolov}, {Guziy}, {Oates}, {Jeong}, {Pandey},
  {Carrasco}, \& {Reverte-Paya}}]{Castro-Tirado2019}
{Castro-Tirado}, A.~J., {Hu}, Y., {Fernandez-Garcia}, E., {et~al.} 2019, GRB
  Coordinates Network, 23708, 1

\bibitem[{{Chevalier} \& {Li}(1999)}]{Chevalier1999}
{Chevalier}, R.~A., \& {Li}, Z.-Y. 1999, \apjl, 520, L29

\bibitem[{{Dai} \& {Lu}(1998)}]{Dai1998MNRAS}
{Dai}, Z.~G., \& {Lu}, T. 1998, \mnras, 298, 87

\bibitem[{{de Ugarte Postigo} {et~al.}(2020){de Ugarte Postigo}, {Th{\"o}ne},
  {Mart{\'\i}n}, {Japelj}, {Levan}, {Micha{\l}owski}, {Selsing}, {Kann},
  {Schulze}, {Palmerio}, {Vergani}, {Tanvir}, {Bensch}, {Covino}, {D'Elia}, {De
  Pasquale}, {Fruchter}, {Fynbo}, {Hartmann}, {Heintz}, {van der Horst},
  {Izzo}, {Jakobsson}, {Ng}, {Perley}, {Rossi}, {Sbarufatti}, {Salvaterra},
  {S{\'a}nchez-Ram{\'\i}rez}, {Watson}, \& {Xu}}]{deUgartePostigo2020}
{de Ugarte Postigo}, A., {Th{\"o}ne}, C.~C., {Mart{\'\i}n}, S., {et~al.} 2020,
  \aap, 633, A68

\bibitem[{{Evans} {et~al.}(2007){Evans}, {Beardmore}, {Page}, {Tyler},
  {Osborne}, {Goad}, {O'Brien}, {Vetere}, {Racusin}, {Morris}, {Burrows},
  {Capalbi}, {Perri}, {Gehrels}, \& {Romano}}]{Evans2007}
{Evans}, P.~A., {Beardmore}, A.~P., {Page}, K.~L., {et~al.} 2007, \aap, 469,
  379

\bibitem[{{Evans} {et~al.}(2009){Evans}, {Beardmore}, {Page}, {Osborne},
  {O'Brien}, {Willingale}, {Starling}, {Burrows}, {Godet}, {Vetere}, {Racusin},
  {Goad}, {Wiersema}, {Angelini}, {Capalbi}, {Chincarini}, {Gehrels}, {Kennea},
  {Margutti}, {Morris}, {Mountford}, {Pagani}, {Perri}, {Romano}, \&
  {Tanvir}}]{Evans2009}
---. 2009, {Monthly Notices of the Royal Astronomical Society}, 397, 1177

\bibitem[{{Foreman-Mackey} {et~al.}(2013){Foreman-Mackey}, {Hogg}, {Lang}, \&
  {Goodman}}]{Foreman-Mackey2013}
{Foreman-Mackey}, D., {Hogg}, D.~W., {Lang}, D., \& {Goodman}, J. 2013, \pasp,
  125, 306

\bibitem[{{Gao} \& {Zhang}(2015)}]{Gao2015}
{Gao}, H., \& {Zhang}, B. 2015, {The Astrophysical Journal}, 801, 103

\bibitem[{{Ghirlanda} {et~al.}(2007){Ghirlanda}, {Bosnjak}, {Ghisellini},
  {Tavecchio}, \& {Firmani}}]{Ghirlanda2007}
{Ghirlanda}, G., {Bosnjak}, Z., {Ghisellini}, G., {Tavecchio}, F., \&
  {Firmani}, C. 2007, \mnras, 379, 73

\bibitem[{{Goldstein} {et~al.}(2012){Goldstein}, {Burgess}, {Preece}, {Briggs},
  {Guiriec}, {van der Horst}, {Connaughton}, {Wilson-Hodge}, {Paciesas},
  {Meegan}, {von Kienlin}, {Bhat}, {Bissaldi}, {Chaplin}, {Diehl}, {Fishman},
  {Fitzpatrick}, {Foley}, {Gibby}, {Giles}, {Greiner}, {Gruber}, {Kippen},
  {Kouveliotou}, {McBreen}, {McGlynn}, {Rau}, \& {Tierney}}]{Goldstein2012}
{Goldstein}, A., {Burgess}, J.~M., {Preece}, R.~D., {et~al.} 2012, {The
  Astrophysical Journal}s, 199, 19

\bibitem[{{Goodman}(1986)}]{Goodman1986}
{Goodman}, J. 1986, \apjl, 308, L47

\bibitem[{{Guiriec} {et~al.}(2011){Guiriec}, {Connaughton}, {Briggs},
  {Burgess}, {Ryde}, {Daigne}, {M{\'e}sz{\'a}ros}, {Goldstein}, {McEnery},
  {Omodei}, {Bhat}, {Bissaldi}, {Camero-Arranz}, {Chaplin}, {Diehl}, {Fishman},
  {Foley}, {Gibby}, {Giles}, {Greiner}, {Gruber}, {von Kienlin}, {Kippen},
  {Kouveliotou}, {McBreen}, {Meegan}, {Paciesas}, {Preece}, {Rau}, {Tierney},
  {van der Horst}, \& {Wilson-Hodge}}]{Guiriec2011}
{Guiriec}, S., {Connaughton}, V., {Briggs}, M.~S., {et~al.} 2011, {The
  Astrophysical Journal Letters}, 727, L33

\bibitem[{{Guiriec} {et~al.}(2013){Guiriec}, {Daigne}, {Hasco{\"e}t},
  {Vianello}, {Ryde}, {Mochkovitch}, {Kouveliotou}, {Xiong}, {Bhat}, {Foley},
  {Gruber}, {Burgess}, {McGlynn}, {McEnery}, \& {Gehrels}}]{Guiriec2013}
{Guiriec}, S., {Daigne}, F., {Hasco{\"e}t}, R., {et~al.} 2013, {The
  Astrophysical Journal}, 770, 32

\bibitem[{{Hamburg} {et~al.}(2019){Hamburg}, {Veres}, {Meegan}, {Burns},
  {Connaughton}, {Goldstein}, {Kocevski}, \& {Roberts}}]{Hamburg2019}
{Hamburg}, R., {Veres}, P., {Meegan}, C., {et~al.} 2019, GRB Coordinates
  Network, Circular Service, No.~23707, \#1 (2019), 23707

\bibitem[{{Im} {et~al.}(2019{\natexlab{a}}){Im}, {Paek}, {Kim}, {Lim}, \&
  {Choi}}]{Im2019}
{Im}, M., {Paek}, G.~S., {Kim}, S., {Lim}, G., \& {Choi}, C.
  2019{\natexlab{a}}, GRB Coordinates Network, 23717, 1

\bibitem[{{Im} {et~al.}(2019{\natexlab{b}}){Im}, {Paek}, \& {Choi}}]{Im2019a}
{Im}, M., {Paek}, G.~S.~H., \& {Choi}, C. 2019{\natexlab{b}}, GRB Coordinates
  Network, 23757, 1

\bibitem[{{J.D. Gropp} {et~al.}(2019){J.D. Gropp}, Kennea, Krimm, LaPorte,
  Lien, Moss, D.~M.~Palmer, \& Siegel}]{J.D.Gropp2019}
{J.D. Gropp}, Kennea, J.~A., Krimm, N. J. K. P. H.~A., {et~al.} 2019, GRB
  Coordinates Network

\bibitem[{{Jordana-Mitjans} {et~al.}(2020){Jordana-Mitjans}, {Mundell},
  {Kobayashi}, {Smith}, {Guidorzi}, {Steele}, {Shrestha}, {Gomboc}, {Marongiu},
  {Martone}, {Lipunov}, {Gorbovskoy}, {Buckley}, {Rebolo}, \&
  {Budnev}}]{Jordana-Mitjans2020}
{Jordana-Mitjans}, N., {Mundell}, C.~G., {Kobayashi}, S., {et~al.} 2020, {The
  Astrophysical Journal}, 892, 97

\bibitem[{{Kim} \& {Im}(2019)}]{Kim2019a}
{Kim}, J., \& {Im}, M. 2019, GRB Coordinates Network, 23732, 1

\bibitem[{{Kim} {et~al.}(2019){Kim}, {Im}, {Lee}, {Kim}, {de Ugrate Postigo},
  \& {Castro-Tirado}}]{Kim2019}
{Kim}, J., {Im}, M., {Lee}, C.~U., {et~al.} 2019, GRB Coordinates Network,
  23734, 1

\bibitem[{{Kobayashi} \& {Sari}(2001)}]{Kobayashi2001}
{Kobayashi}, S., \& {Sari}, R. 2001, \apj, 551, 934

\bibitem[{{Kobayashi} \& {Zhang}(2003)}]{Kobayashi2003}
{Kobayashi}, S., \& {Zhang}, B. 2003, \apj, 597, 455

\bibitem[{{Kocevski} {et~al.}(2019){Kocevski}, {Omodei}, {Axelsson}, {Burns},
  {Vianello}, {Bissaldi}, \& {Longo}}]{Kocevski2019}
{Kocevski}, D., {Omodei}, N., {Axelsson}, M., {et~al.} 2019, GRB Coordinates
  Network, Circular Service, No.~23709, \#1 (2019/January-0), 23709

\bibitem[{{Laskar} {et~al.}(2019){Laskar}, {Alexander}, {Gill}, {Granot},
  {Berger}, {Mundell}, {Barniol Duran}, {Bolmer}, {Duffell}, {van Eerten},
  {Fong}, {Kobayashi}, {Margutti}, \& {Schady}}]{Laskar2019}
{Laskar}, T., {Alexander}, K.~D., {Gill}, R., {et~al.} 2019, \apjl, 878, L26

\bibitem[{{Li}(2019{\natexlab{a}})}]{Li2019a}
{Li}, L. 2019{\natexlab{a}}, {The Astrophysical Journal}s, 242, 16

\bibitem[{{Li}(2019{\natexlab{b}})}]{Li2019c}
---. 2019{\natexlab{b}}, \apjs, 245, 7

\bibitem[{{Li} \& {Zhang}(2021)}]{Li2021a}
{Li}, L., \& {Zhang}, B. 2021, \apjs, 253, 43

\bibitem[{{Liang} {et~al.}(2010){Liang}, {Yi}, {Zhang}, {L{\"u}}, {Zhang}, \&
  {Zhang}}]{Liang2010}
{Liang}, E.-W., {Yi}, S.-X., {Zhang}, J., {et~al.} 2010, \apj, 725, 2209

\bibitem[{{Liang} {et~al.}(2013){Liang}, {Li}, {Gao}, {Zhang}, {Liang}, {Wu},
  {Yi}, {Dai}, {Tang}, {Chen}, {L{\"u}}, {Zhang}, {Lu}, {L{\"u}}, \&
  {Wei}}]{Liang2013}
{Liang}, E.-W., {Li}, L., {Gao}, H., {et~al.} 2013, \apj, 774, 13

\bibitem[{{Lloyd-Ronning} \& {Zhang}(2004)}]{Lloyd-Ronning2004}
{Lloyd-Ronning}, N.~M., \& {Zhang}, B. 2004, {The Astrophysical Journal}, 613,
  477

\bibitem[{{MAGIC Collaboration} {et~al.}(2019{\natexlab{a}}){MAGIC
  Collaboration}, {Acciari}, {Ansoldi}, {Antonelli}, {Engels}, {Baack},
  {Babi{\'c}}, {Banerjee}, {Barres de Almeida}, {Barrio}, {Becerra
  Gonz{\'a}lez}, {Bednarek}, {Bellizzi}, {Bernardini}, {Berti}, {Besenrieder},
  {Bhattacharyya}, {Bigongiari}, {Biland }, {Blanch}, {Bonnoli},
  {Bo{\v{s}}njak}, {Busetto}, {Carosi}, {Ceribella}, {Chai}, {Chilingaryan},
  {Cikota}, {Colak}, {Colin}, {Colombo}, {Contreras}, {Cortina}, {Covino},
  {D'Elia}, {da Vela}, {Dazzi}, {de Angelis}, {de Lotto}, {Delfino}, {Delgado},
  {Depaoli}, {di Pierro}, {di Venere}, {Do Souto Espi{\~n}eira}, {Dominis
  Prester}, {Donini}, {Dorner}, {Doro}, {Elsaesser}, {Fallah Ramazani},
  {Fattorini}, {Ferrara}, {Fidalgo}, {Foffano}, {Fonseca}, {Font}, {Fruck},
  {Fukami}, {Garc{\'\i}a L{\'o}pez}, {Garczarczyk}, {Gasparyan}, {Gaug},
  {Giglietto}, {Giordano}, {Godinovi{\'c}}, {Green}, {Guberman}, {Hadasch},
  {Hahn}, {Herrera}, {Hoang}, {Hrupec}, {H{\"u}tten}, {Inada}, {Inoue},
  {Ishio}, {Iwamura}, {Jouvin}, {Kerszberg}, {Kubo}, {Kushida}, {Lamastra},
  {Lelas}, {Leone}, {Lindfors}, {Lombardi}, {Longo}, {L{\'o}pez},
  {L{\'o}pez-Coto}, {L{\'o}pez-Oramas}, {Loporchio}, {Machado de Oliveira
  Fraga}, {Maggio}, {Majumdar}, {Makariev}, {Mallamaci}, {Maneva}, {Manganaro},
  {Mannheim}, {Maraschi}, {Mariotti}, {Mart{\'\i}nez}, {Mazin},
  {Mi{\'c}anovi{\'c}}, {Miceli}, {Minev}, {Mirand a}, {Mirzoyan}, {Molina},
  {Moralejo}, {Morcuende}, {Moreno}, {Moretti}, {Munar-Adrover}, {Neustroev},
  {Nigro}, {Nilsson}, {Ninci}, {Nishijima}, {Noda}, {Nogu{\'e}s}, {Nozaki},
  {Paiano}, {Palatiello}, {Paneque}, {Paoletti}, {Paredes}, {Pe{\~n}il},
  {Peresano}, {Persic}, {Moroni}, {Prandini}, {Puljak}, {Rhode}, {Rib{\'o}},
  {Rico}, {Righi}, {Rugliancich}, {Saha}, {Sahakyan}, {Saito}, {Sakurai},
  {Satalecka}, {Schmidt}, {Schweizer}, {Sitarek}, {{\v{S}}nidari{\'c}},
  {Sobczynska}, {Somero}, {Stamerra}, {Strom}, {Strzys}, {Suda}, {Suri{\'c}},
  {Takahashi}, {Tavecchio}, {Temnikov}, {Terzi{\'c}}, {Teshima},
  {Torres-Alb{\`a}}, {Tosti}, {Vagelli}, {van Scherpenberg}, {Vanzo}, {Vazquez
  Acosta}, {Vigorito}, {Vitale}, {Vovk}, {Will}, {Zari{\'c}}, {Nava}, {Veres},
  {Bhat}, {Briggs}, {Cleveland }, {Hamburg}, {Hui}, {Mailyan}, {Preece},
  {Roberts}, {von Kienlin}, {Wilson-Hodge}, {Kocevski}, {Arimoto}, {Tak},
  {Asano}, {Axelsson}, {Barbiellini}, {Bissaldi}, {Dirirsa}, {Gill}, {Granot},
  {McEnery}, {Omodei}, {Razzaque}, {Piron}, {Racusin}, {Thompson}, {Campana},
  {Bernardini}, {Kuin}, {Siegel}, {Cenko}, {O'Brien}, {Capalbi}, {Da{\i}}, {de
  Pasquale}, {Gropp}, {Klingler}, {Osborne}, {Perri}, {Starling},
  {Tagliaferri}, {Tohuvavohu}, {Ursi}, {Tavani}, {Cardillo}, {Casentini},
  {Piano}, {Evangelista}, {Verrecchia}, {Pittori}, {Lucarelli}, {Bulgarelli},
  {Parmiggiani}, {Anderson}, {Anderson}, {Bernardi}, {Bolmer},
  {Caballero-Garc{\'\i}a}, {Carrasco}, {Castell{\'o}n}, {Castro Segura},
  {Castro-Tirado}, {Cherukuri}, {Cockeram}, {D'Avanzo}, {di Dato}, {Diretse},
  {Fender}, {Fern{\'a}ndez-Garc{\'\i}a}, {Fynbo}, {Fruchter}, {Greiner},
  {Gromadzki}, {Heintz}, {Heywood}, {van der Horst}, {Hu}, {Inserra}, {Izzo},
  {Jaiswal}, {Jakobsson}, {Japelj}, {Kankare}, {Kann}, {Kouveliotou}, {Klose},
  {Levan}, {Li}, {Lotti}, {Maguire}, {Malesani}, {Manulis}, {Marongiu},
  {Martin}, {Melandri}, {Micha{\l}owski}, {Miller-Jones}, {Misra}, {Moin},
  {Mooley}, {Nasri}, {Nicholl}, {Noschese}, {Novara}, {Pandey}, {Peretti},
  {P{\'e}rez Del Pulgar}, {P{\'e}rez-Torres}, {Perley}, {Piro}, {Ragosta},
  {Resmi}, {Ricci}, {Rossi}, {S{\'a}nchez-Ram{\'\i}rez}, {Selsing}, {Schulze},
  {Smartt}, {Smith}, {Sokolov}, {Stevens}, {Tanvir}, {Th{\"o}ne}, {Tiengo},
  {Tremou}, {Troja}, {de Ugarte Postigo}, {Valeev}, {Vergani}, {Wieringa},
  {Woudt}, {Xu}, {Yaron}, \& {Young}}]{MAGICCollaboration2019b}
{MAGIC Collaboration}, {Acciari}, V.~A., {Ansoldi}, S., {et~al.}
  2019{\natexlab{a}}, \nat, 575, 459

\bibitem[{{MAGIC Collaboration} {et~al.}(2019{\natexlab{b}}){MAGIC
  Collaboration}, {Acciari}, {Ansoldi}, {Antonelli}, {Arbet Engels}, {Baack},
  {Babi{\'c}}, {Banerjee}, {Barres de Almeida}, {Barrio}, {Becerra
  Gonz{\'a}lez}, {Bednarek}, {Bellizzi}, {Bernardini}, {Berti}, {Besenrieder},
  {Bhattacharyya}, {Bigongiari}, {Biland }, {Blanch}, {Bonnoli},
  {Bo{\v{s}}njak}, {Busetto}, {Carosi}, {Carosi}, {Ceribella}, {Chai},
  {Chilingaryan}, {Cikota}, {Colak}, {Colin}, {Colombo}, {Contreras},
  {Cortina}, {Covino}, {D'Amico}, {D'Elia}, {da Vela}, {Dazzi}, {de Angelis},
  {de Lotto}, {Delfino}, {Delgado}, {Depaoli}, {di Pierro}, {di Venere}, {Do
  Souto Espi{\~n}eira}, {Dominis Prester}, {Donini}, {Dorner}, {Doro},
  {Elsaesser}, {Fallah Ramazani}, {Fattorini}, {Fern{\'a}ndez-Barral},
  {Ferrara}, {Fidalgo}, {Foffano}, {Fonseca}, {Font}, {Fruck}, {Fukami},
  {Gallozzi}, {Garc{\'\i}a L{\'o}pez}, {Garczarczyk}, {Gasparyan}, {Gaug},
  {Giglietto}, {Giordano}, {Godinovi{\'c}}, {Green}, {Guberman}, {Hadasch},
  {Hahn}, {Herrera}, {Hoang}, {Hrupec}, {H{\"u}tten}, {Inada}, {Inoue},
  {Ishio}, {Iwamura}, {Jouvin}, {Kerszberg}, {Kubo}, {Kushida}, {Lamastra},
  {Lelas}, {Leone}, {Lindfors}, {Lombardi}, {Longo}, {L{\'o}pez},
  {L{\'o}pez-Coto}, {L{\'o}pez-Oramas}, {Loporchio}, {Machado de Oliveira
  Fraga}, {Maggio}, {Majumdar}, {Makariev}, {Mallamaci}, {Maneva}, {Manganaro},
  {Mannheim}, {Maraschi}, {Mariotti}, {Mart{\'\i}nez}, {Masuda}, {Mazin},
  {Mi{\'c}anovi{\'c}}, {Miceli}, {Minev}, {Miranda}, {Mirzoyan}, {Molina},
  {Moralejo}, {Morcuende}, {Moreno}, {Moretti}, {Munar-Adrover}, {Neustroev},
  {Nigro}, {Nilsson}, {Ninci}, {Nishijima}, {Noda}, {Nogu{\'e}s}, {N{\"o}the},
  {Nozaki}, {Paiano}, {Palacio}, {Palatiello}, {Paneque}, {Paoletti},
  {Paredes}, {Pe{\~n}il}, {Peresano}, {Persic}, {Prada Moroni}, {Prand ini},
  {Puljak}, {Rhode}, {Rib{\'o}}, {Rico}, {Righi}, {Rugliancich}, {Saha},
  {Sahakyan}, {Saito}, {Sakurai}, {Satalecka}, {Schmidt}, {Schweizer},
  {Sitarek}, {{\v{S}}nidari{\'c}}, {Sobczynska}, {Somero}, {Stamerra}, {Strom},
  {Strzys}, {Suda}, {Suri{\'c}}, {Takahashi}, {Tavecchio}, {Temnikov},
  {Terzi{\'c}}, {Teshima}, {Torres-Alb{\`a}}, {Tosti}, {Tsujimoto}, {Vagelli},
  {van Scherpenberg}, {Vanzo}, {Vazquez Acosta}, {Vigorito}, {Vitale}, {Vovk},
  {Will}, {Zari{\'c}}, \& {Nava}}]{MAGICCollaboration2019}
---. 2019{\natexlab{b}}, \nat, 575, 455

\bibitem[{{Mazaeva} {et~al.}(2019){Mazaeva}, {Pozanenko}, {Volnova}, {Belkin},
  \& {Krugov}}]{Mazaeva2019}
{Mazaeva}, E., {Pozanenko}, A., {Volnova}, A., {Belkin}, S., \& {Krugov}, M.
  2019, GRB Coordinates Network, 23741, 1

\bibitem[{{Melandri} {et~al.}(2021){Melandri}, {Izzo}, {Pian}, {Malesani},
  {Della Valle}, {Rossi}, {D'Avanzo}, {Guetta}, {Mazzali}, {Benetti},
  {Masetti}, {Palazzi}, {Savaglio}, {Amati}, {Antonelli}, {Ashall},
  {Bernardini}, {Campana}, {Carini}, {Covino}, {D'Elia}, {de Ugarte Postigo},
  {De Pasquale}, {Filippenko}, {Fruchter}, {Fynbo}, {Giunta}, {Hartmann},
  {Jakobsson}, {Japelj}, {Jonker}, {Kann}, {Lamb}, {Levan}, {Martin-Carrillo},
  {Moller}, {Piranomonte}, {Pugliese}, {Salvaterra}, {Schulze}, {Starling},
  {Stella}, {Tagliaferri}, {Tanvir}, \& {Watson}}]{2021arXiv211204759M}
{Melandri}, A., {Izzo}, L., {Pian}, E., {et~al.} 2021, arXiv e-prints,
  arXiv:2112.04759

\bibitem[{{Meszaros} \& {Rees}(1993)}]{Meszaros1993}
{Meszaros}, P., \& {Rees}, M.~J. 1993, {The Astrophysical Journal Letters},
  418, L59

\bibitem[{{M{\'e}sz{\'a}ros} \& {Rees}(2000)}]{Meszaros2000}
{M{\'e}sz{\'a}ros}, P., \& {Rees}, M.~J. 2000, {The Astrophysical Journal},
  530, 292

\bibitem[{{M{\'e}sz{\'a}ros} {et~al.}(1998){M{\'e}sz{\'a}ros}, {Rees}, \&
  {Wijers}}]{Meszaros1998}
{M{\'e}sz{\'a}ros}, P., {Rees}, M.~J., \& {Wijers}, R.~A.~M.~J. 1998, ApJ, 499,
  301

\bibitem[{{Misra} {et~al.}(2019){Misra}, {Resmi}, {Kann}, {Marongiu}, {Moin},
  {Klose}, {de Ugarte Postigo}, {Jaiswal}, {Perley}, {Ghosh}, {Bernardi},
  {Schulze}, {Micha{\l}owski}, {Mart{\'\i}n}, {Cockeram}, {Kumar}, {Cherukuri},
  {Bhalerao}, {Anderson}, {Anupama}, {Th{\"o}ne}, {Barway}, {Wieringa},
  {Fynbo}, \& {Habeeb}}]{Misra2019}
{Misra}, K., {Resmi}, L., {Kann}, D.~A., {et~al.} 2019, {Monthly Notices of the
  Royal Astronomical Society}, submitted (arXiv:1911.09719), arXiv:1911.09719

\bibitem[{{Narayana Bhat} {et~al.}(2016){Narayana Bhat}, {Meegan}, {von
  Kienlin}, {Paciesas}, {Briggs}, {Burgess}, {Burns}, {Chaplin}, {Cleveland},
  {Collazzi}, {Connaughton}, {Diekmann}, {Fitzpatrick}, {Gibby}, {Giles},
  {Goldstein}, {Greiner}, {Jenke}, {Kippen}, {Kouveliotou}, {Mailyan},
  {McBreen}, {Pelassa}, {Preece}, {Roberts}, {Sparke}, {Stanbro}, {Veres},
  {Wilson-Hodge}, {Xiong}, {Younes}, {Yu}, \& {Zhang}}]{NarayanaBhat2016}
{Narayana Bhat}, P., {Meegan}, C.~A., {von Kienlin}, A., {et~al.} 2016, {The
  Astrophysical Journal}s, 223, 28

\bibitem[{{Paczynski}(1986)}]{Paczynski1986}
{Paczynski}, B. 1986, \apjl, 308, L43

\bibitem[{{Panaitescu} \& {Kumar}(2001)}]{Panaitescu2001}
{Panaitescu}, A., \& {Kumar}, P. 2001, \apjl, 560, L49

\bibitem[{{Panaitescu} \& {Kumar}(2002)}]{Panaitescu2002}
---. 2002, {The Astrophysical Journal}, 571, 779

\bibitem[{{Pe'er}(2015)}]{Peer2015}
{Pe'er}, A. 2015, Advances in Astronomy, 2015, 907321

\bibitem[{{Pe'er} {et~al.}(2007){Pe'er}, {Ryde}, {Wijers}, {M{\'e}sz{\'a}ros},
  \& {Rees}}]{Peer2007}
{Pe'er}, A., {Ryde}, F., {Wijers}, R.~A.~M.~J., {M{\'e}sz{\'a}ros}, P., \&
  {Rees}, M.~J. 2007, {The Astrophysical Journal Letters}, 664, L1

\bibitem[{{Planck Collaboration} {et~al.}(2018){Planck Collaboration},
  {Aghanim}, {Akrami}, {Ashdown}, {Aumont}, {Baccigalupi}, {Ballardini},
  {Banday}, {Barreiro}, {Bartolo}, {Basak}, {Battye}, {Benabed}, {Bernard},
  {Bersanelli}, {Bielewicz}, {Bock}, {Bond}, {Borrill}, {Bouchet}, {Boulanger},
  {Bucher}, {Burigana}, {Butler}, {Calabrese}, {Cardoso}, {Carron},
  {Challinor}, {Chiang}, {Chluba}, {Colombo}, {Combet}, {Contreras}, {Crill},
  {Cuttaia}, {de Bernardis}, {de Zotti}, {Delabrouille}, {Delouis}, {Di
  Valentino}, {Diego}, {Dor{\'e}}, {Douspis}, {Ducout}, {Dupac}, {Dusini},
  {Efstathiou}, {Elsner}, {En{\ss}lin}, {Eriksen}, {Fantaye}, {Farhang},
  {Fergusson}, {Fernandez-Cobos}, {Finelli}, {Forastieri}, {Frailis},
  {Franceschi}, {Frolov}, {Galeotta}, {Galli}, {Ganga}, {G{\'e}nova-Santos},
  {Gerbino}, {Ghosh}, {Gonz{\'a}lez-Nuevo}, {G{\'o}rski}, {Gratton},
  {Gruppuso}, {Gudmundsson}, {Hamann}, {Hand ley}, {Herranz}, {Hivon}, {Huang},
  {Jaffe}, {Jones}, {Karakci}, {Keih{\"a}nen}, {Keskitalo}, {Kiiveri}, {Kim},
  {Kisner}, {Knox}, {Krachmalnicoff}, {Kunz}, {Kurki-Suonio}, {Lagache},
  {Lamarre}, {Lasenby}, {Lattanzi}, {Lawrence}, {Le Jeune}, {Lemos},
  {Lesgourgues}, {Levrier}, {Lewis}, {Liguori}, {Lilje}, {Lilley}, {Lindholm},
  {L{\'o}pez-Caniego}, {Lubin}, {Ma}, {Mac{\'\i}as-P{\'e}rez}, {Maggio},
  {Maino}, {Mandolesi}, {Mangilli}, {Marcos-Caballero}, {Maris}, {Martin},
  {Martinelli}, {Mart{\'\i}nez-Gonz{\'a}lez}, {Matarrese}, {Mauri}, {McEwen},
  {Meinhold}, {Melchiorri}, {Mennella}, {Migliaccio}, {Millea}, {Mitra},
  {Miville-Desch{\^e}nes}, {Molinari}, {Montier}, {Morgante}, {Moss}, {Natoli},
  {N{\o}rgaard-Nielsen}, {Pagano}, {Paoletti}, {Partridge}, {Patanchon},
  {Peiris}, {Perrotta}, {Pettorino}, {Piacentini}, {Polastri}, {Polenta},
  {Puget}, {Rachen}, {Reinecke}, {Remazeilles}, {Renzi}, {Rocha}, {Rosset},
  {Roudier}, {Rubi{\~n}o-Mart{\'\i}n}, {Ruiz-Granados}, {Salvati}, {Sandri},
  {Savelainen}, {Scott}, {Shellard}, {Sirignano}, {Sirri}, {Spencer},
  {Sunyaev}, {Suur-Uski}, {Tauber}, {Tavagnacco}, {Tenti}, {Toffolatti},
  {Tomasi}, {Trombetti}, {Valenziano}, {Valiviita}, {Van Tent}, {Vibert},
  {Vielva}, {Villa}, {Vittorio}, {Wand elt}, {Wehus}, {White}, {White},
  {Zacchei}, \& {Zonca}}]{PlanckCollaboration2018}
{Planck Collaboration}, {Aghanim}, N., {Akrami}, Y., {et~al.} 2018, arXiv
  e-prints, arXiv:1807.06209

\bibitem[{{Preece} {et~al.}(2016){Preece}, {Goldstein}, {Bhat}, {Stanbro},
  {Hakkila}, \& {Blalock}}]{Preece2016}
{Preece}, R., {Goldstein}, A., {Bhat}, N., {et~al.} 2016, {The Astrophysical
  Journal}, 821, 12

\bibitem[{{Preece} {et~al.}(1998){Preece}, {Briggs}, {Mallozzi}, {Pendleton},
  {Paciesas}, \& {Band}}]{Preece1998}
{Preece}, R.~D., {Briggs}, M.~S., {Mallozzi}, R.~S., {et~al.} 1998, {The
  Astrophysical Journal Letters}, 506, L23

\bibitem[{{Ravasio} {et~al.}(2019){Ravasio}, {Oganesyan}, {Salafia},
  {Ghirlanda}, {Ghisellini}, {Branchesi}, {Campana}, {Covino}, \&
  {Salvaterra}}]{2019A&A...626A..12R}
{Ravasio}, M.~E., {Oganesyan}, G., {Salafia}, O.~S., {et~al.} 2019, \aap, 626,
  A12

\bibitem[{{Rees} \& {Meszaros}(1994)}]{Rees1994}
{Rees}, M.~J., \& {Meszaros}, P. 1994, \apjl, 430, L93

\bibitem[{{Ryde}(2004)}]{Ryde2004}
{Ryde}, F. 2004, {The Astrophysical Journal}, 614, 827

\bibitem[{{Ryde}(2005)}]{Ryde2005}
---. 2005, {The Astrophysical Journal Letters}, 625, L95

\bibitem[{{Ryde} \& {Pe'er}(2009)}]{Ryde2009}
{Ryde}, F., \& {Pe'er}, A. 2009, {The Astrophysical Journal}, 702, 1211

\bibitem[{{Ryde} {et~al.}(2010){Ryde}, {Axelsson}, {Zhang}, {McGlynn}, {Pe'er},
  {Lundman}, {Larsson}, {Battelino}, {Zhang}, {Bissaldi}, {Bregeon}, {Briggs},
  {Chiang}, {de Palma}, {Guiriec}, {Larsson}, {Longo}, {McBreen}, {Omodei},
  {Petrosian}, {Preece}, \& {van der Horst}}]{Ryde2010}
{Ryde}, F., {Axelsson}, M., {Zhang}, B.~B., {et~al.} 2010, {The Astrophysical
  Journal Letters}, 709, L172

\bibitem[{{Ryde} {et~al.}(2011){Ryde}, {Pe'er}, {Nymark}, {Axelsson},
  {Moretti}, {Lundman}, {Battelino}, {Bissaldi}, {Chiang}, {Jackson},
  {Larsson}, {Longo}, {McGlynn}, \& {Omodei}}]{Ryde2011}
{Ryde}, F., {Pe'er}, A., {Nymark}, T., {et~al.} 2011, {Monthly Notices of the
  Royal Astronomical Society}, 415, 3693

\bibitem[{{Sari} \& {Piran}(1999)}]{Sari1999b}
{Sari}, R., \& {Piran}, T. 1999, \apj, 520, 641

\bibitem[{{Sari} {et~al.}(1998){Sari}, {Piran}, \& {Narayan}}]{Sari1998}
{Sari}, R., {Piran}, T., \& {Narayan}, R. 1998, {The Astrophysical Journal
  Letters}, 497, L17

\bibitem[{{Scargle} {et~al.}(2013){Scargle}, {Norris}, {Jackson}, \&
  {Chiang}}]{Scargle2013}
{Scargle}, J.~D., {Norris}, J.~P., {Jackson}, B., \& {Chiang}, J. 2013, {The
  Astrophysical Journal}, 764, 167

\bibitem[{{Schulze} {et~al.}(2011){Schulze}, {Klose}, {Bj{\"o}rnsson},
  {Jakobsson}, {Kann}, {Rossi}, {Kr{\"u}hler}, {Greiner}, \&
  {Ferrero}}]{Schulze2011}
{Schulze}, S., {Klose}, S., {Bj{\"o}rnsson}, G., {et~al.} 2011, \aap, 526, A23

\bibitem[{{Selsing} {et~al.}(2019){Selsing}, {Fynbo}, {Heintz}, \&
  {Watson}}]{Selsing2019}
{Selsing}, J., {Fynbo}, J.~P.~U., {Heintz}, K.~E., \& {Watson}, D. 2019, GRB
  Coordinates Network, 23695, 1

\bibitem[{{Tremou} {et~al.}(2019){Tremou}, {Heywood}, {Vergani}, {Woudt},
  {Fender}, {Horesh}, {Passmoor}, \& {Goedhart}}]{Tremou2019}
{Tremou}, L., {Heywood}, I., {Vergani}, S.~D., {et~al.} 2019, GRB Coordinates
  Network, 23760, 1

\bibitem[{{Ursi} {et~al.}(2020){Ursi}, {Tavani}, {Frederiks}, {Romani},
  {Verrecchia}, {Marisaldi}, {Aptekar}, {Antonelli}, {Argan}, {Bulgarelli},
  {Barbiellini}, {Caraveo}, {Cardillo}, {Casentini}, {Cattaneo}, {Chen},
  {Costa}, {Donnarumma}, {Evangelista}, {Feroci}, {Ferrari}, {Fuschino},
  {Galli}, {Giuliani}, {Labanti}, {Lazzarotto}, {Longo}, {Lucarelli},
  {Morselli}, {Paoletti}, {Parmiggiani}, {Piano}, {Pilia}, {Pittori},
  {Svinkin}, {Trois}, {Tsvetkova}, {Vercellone}, \& {Vittorini}}]{Ursi2020}
{Ursi}, A., {Tavani}, M., {Frederiks}, D.~D., {et~al.} 2020, \apj, 904, 133

\bibitem[{Vereshchagin \& Aksenov(2017)}]{vereshchagin2017}
Vereshchagin, G.~V., \& Aksenov, A.~G. 2017, Relativistic kinetic theory: with
  applications in astrophysics and cosmology (Cambridge University Press)

\bibitem[{{Vianello} {et~al.}(2015){Vianello}, {Lauer}, {Younk}, {Tibaldo},
  {Burgess}, {Ayala}, {Harding}, {Hui}, {Omodei}, \& {Zhou}}]{Vianello2015}
{Vianello}, G., {Lauer}, R.~J., {Younk}, P., {et~al.} 2015, arXiv e-prints,
  arXiv:1507.08343

\bibitem[{Vuong(1989)}]{Vuong1989}
Vuong, Q.~H. 1989, Econometrica, 57, 307

\bibitem[{{Wang} {et~al.}(2019{\natexlab{a}}){Wang}, {Liu}, {Zhang}, {Xi}, \&
  {Zhang}}]{Wang2019b}
{Wang}, X.-Y., {Liu}, R.-Y., {Zhang}, H.-M., {Xi}, S.-Q., \& {Zhang}, B.
  2019{\natexlab{a}}, {The Astrophysical Journal}, 884, 117

\bibitem[{{Wang} {et~al.}(2019{\natexlab{b}}){Wang}, {Li}, {Moradi}, \&
  {Ruffini}}]{Wang2019}
{Wang}, Y., {Li}, L., {Moradi}, R., \& {Ruffini}, R. 2019{\natexlab{b}}, arXiv
  e-prints, arXiv:1901.07505

\bibitem[{{Watson} {et~al.}(2019{\natexlab{a}}){Watson}, {Butler}, {Becerra},
  {Gonzalez}, {Lee}, {Pereyra}, {Roman-Zuniga}, {Kutyrev}, \&
  {Troja}}]{Watson2019}
{Watson}, A.~M., {Butler}, N., {Becerra}, R.~L., {et~al.} 2019{\natexlab{a}},
  GRB Coordinates Network, 23749, 1

\bibitem[{{Watson} {et~al.}(2019{\natexlab{b}}){Watson}, {Butler}, {Kutyrev},
  {Lee}, {Richer}, {Fox}, {Prochaska}, {Bloom}, {Cucchiara}, {Troja},
  {Littlejohns}, {Ramirez-Ruiz}, {Gonzalez}, {Roman-Zuniga}, {Moseley},
  {Capone}, {Golkhou}, \& {Toy}}]{Watson2019a}
{Watson}, A.~M., {Butler}, N., {Kutyrev}, A., {et~al.} 2019{\natexlab{b}}, GRB
  Coordinates Network, 23751, 1

\bibitem[{{Wu} {et~al.}(2003){Wu}, {Dai}, {Huang}, \& {Lu}}]{Wu2003}
{Wu}, X.~F., {Dai}, Z.~G., {Huang}, Y.~F., \& {Lu}, T. 2003, \mnras, 342, 1131

\bibitem[{{Yost} {et~al.}(2003){Yost}, {Harrison}, {Sari}, \&
  {Frail}}]{Yost2003}
{Yost}, S.~A., {Harrison}, F.~A., {Sari}, R., \& {Frail}, D.~A. 2003, {The
  Astrophysical Journal}, 597, 459

\bibitem[{{Zeh} {et~al.}(2006){Zeh}, {Klose}, \& {Kann}}]{Zeh2006}
{Zeh}, A., {Klose}, S., \& {Kann}, D.~A. 2006, {The Astrophysical Journal},
  637, 889

\bibitem[{{Zhang}(2018)}]{Zhang2018}
{Zhang}, B. 2018, {The Physics of Gamma-Ray Bursts}, doi:10.1017/9781139226530

\bibitem[{{Zhang} {et~al.}(2021){Zhang}, {Wang}, \& {Li}}]{Zhang2021}
{Zhang}, B., {Wang}, Y., \& {Li}, L. 2021, \apjl, 909, L3

\bibitem[{{Zhang} {et~al.}(2007){Zhang}, {Liang}, {Page}, {Grupe}, {Zhang},
  {Barthelmy}, {Burrows}, {Campana}, {Chincarini}, {Gehrels}, {Kobayashi},
  {M{\'e}sz{\'a}ros}, {Moretti}, {Nousek}, {O'Brien}, {Osborne}, {Roming},
  {Sakamoto}, {Schady}, \& {Willingale}}]{Zhang2007b}
{Zhang}, B., {Liang}, E., {Page}, K.~L., {et~al.} 2007, {The Astrophysical
  Journal}, 655, 989

\bibitem[{{Zhang} {et~al.}(2018){Zhang}, {Zhang}, {Castro-Tirado}, {Dai},
  {Tam}, {Wang}, {Hu}, {Karpov}, {Pozanenko}, {Zhang}, {Mazaeva}, {Minaev},
  {Volnova}, {Oates}, {Gao}, {Wu}, {Shao}, {Tang}, {Beskin}, {Biryukov},
  {Bondar}, {Ivanov}, {Katkova}, {Orekhova}, {Perkov}, {Sasyuk}, {Mankiewicz},
  {{\.Z}arnecki}, {Cwiek}, {Opiela}, {Zadro{\.Z}ny}, {Aptekar}, {Frederiks},
  {Svinkin}, {Kusakin}, {Inasaridze}, {Burhonov}, {Rumyantsev}, {Klunko},
  {Moskvitin}, {Fatkhullin}, {Sokolov}, {Valeev}, {Jeong}, {Park},
  {Caballero-Garc{\'{\i}}a}, {Cunniffe}, {Tello}, {Ferrero}, {Pandey},
  {Jel{\'{\i}}nek}, {Peng}, {S{\'a}nchez-Ram{\'{\i}}rez}, \&
  {Castell{\'o}n}}]{Zhang2018a}
{Zhang}, B.-B., {Zhang}, B., {Castro-Tirado}, A.~J., {et~al.} 2018, Nature
  Astronomy, 2, 69

\end{thebibliography}

\clearpage
\begin{table}
\setlength{\tabcolsep}{0.0em}
\renewcommand\arraystretch{1.2}
\scalebox{0.7}
{
\begin{tabular}{cccccccc}
\hline
$t_{1} \sim t_{2}$&$\Delta$DIC(1)&$\Delta$DIC(2)&$\Delta$DIC(3)&$\Delta$DIC(4)&$\Delta$DIC(5)&$\Delta$DIC(6)&$\Delta$DIC(7)\\
(s)&(CPL+BB-PL)&(CPL+BB-BB)&(CPL+BB-CPL)&(CPL+BB-Band)&(CPL+BB-SBKPL)&(CPL+BB)-(PL+BB)&(CPLBB)-(PL+Bandcut)\\
\hline
0$\sim$116&-3523&-19565&-266&-262&-34&-457&-68\\
\hline
\end{tabular}
}
\caption{Comparison of $\Delta$DIC between the best model (CPL+BB) and other various models (PL, BB, CPL, Band, SBKPL, PL+BB, PL+Bandcut) in GRB 190114C, which is based on the time-integrated spectral analysis.}\label{tab:model}
\end{table}

\clearpage
\begin{table*}
\setlength{\tabcolsep}{0.10em}
\renewcommand\arraystretch{1.2}
\centering
\scalebox{0.8}{
\begin{tabular}{c|ccccc}
\hline
Satellite-Instrument&$T_{0}$+[$t_{\rm start}$,$t_{\rm stop}$] 
&Observed Bandwidth&Isotropic Energy&Model&Reference\\
&(s)&&(erg)&(For energy)\\
\hline
MAGIC$^{a}$&$T_{0}$+[62, 2454]&0.3$\sim$1 TeV& $\sim$4.0$\times$10$^{51}$&SPL&\cite{MAGICCollaboration2019}\\
\hline
{\it Fermi}-LAT$^{b}$&$T_{0}$+[2.1, 8000]&0.1$\sim$10 GeV&(1.09$\pm$0.24)$\times$10$^{53}$&SPL&this paper\\
{\it Fermi}-LAT$^{b}$&$T_{0}$+[2.1, 6]&0.1$\sim$10 GeV&(8.49$\pm$1.80)$\times$10$^{51}$&SPL&this paper\\
{\it Fermi}-LAT$^{b}$&$T_{0}$+[6, 8000]&0.1$\sim$10 GeV&(1.01$\pm$0.24)$\times$10$^{53}$&SPL&this paper\\
\hline
{\it Fermi}-GBM$^{c}$&$T_{0}$+[0, 116]&0.001$\sim$10 MeV&(2.82$^{+0.43}_{-0.25}$)$\times$10$^{53}$&(CPL+BB)/Band&this paper\\
{\it Fermi}-GBM$^{c}$&$T_{0}$+[0, 6]&...&(2.29$^{+0.10}_{-0.09}$)$\times$10$^{53}$&CPL+BB&this paper\\
{\it Fermi}-GBM$^{c}$&$T_{0}$+[6, 116]&...&(5.33$^{+4.23}_{-2.34}$)$\times$10$^{52}$&Band&this paper\\
{\it Fermi}-GBM$^{c}$&$T_{0}$+[0, 6]&...&(3.69$^{+0.78}_{-0.67}$)$\times$10$^{52}$&CPL+BB&this paper\\
{\it Fermi}-GBM$^{c}$&$T_{0}$+[0, 6]&...&(1.92$^{+0.12}_{-0.11}$)$\times$10$^{53}$&CPL+BB&this paper\\
\hline
{\it Swift}-XRT$^{d}$&$T_{0}$+[68, 1197626]&0.3$\sim$10 KeV&$\sim$1.48$\times$10$^{52}$&SPL&this paper\\
\hline
\end{tabular}
}
\caption{Various isotropic energy releases were observed by different satellite instruments at different wavelengths and different time intervals. Notes. $^{a}$ Time-integrated-isotropic-equivalent energy releases observed by MAGIC from $T_{0}$+62 to $T_{0}$+2454 s as reported in~\cite{MAGICCollaboration2019}. $^{b}$ Time-integrated-isotropic-equivalent energy releases by the {\it Fermi}-LAT observation with a (0.1-10GeV) bandwidth, as well as separated into the prompt and afterglow emission as defined in the Methods. $^{c}$ Time-integrated-isotropic-equivalent energy releases by the {\it Fermi}-GBM observation using the best models; as well as those separated into the prompt and afterglow emission, and into the thermal and non-thermal energy releases during the prompt emission phase.$^{d}$ The total-isotropic-equivalent energy release observed by the {\it Swift}-XRT.}\label{tab:energy}
\end{table*}

\clearpage
\begin{table}
\setlength{\tabcolsep}{0.10em}
\renewcommand\arraystretch{1.2}
\centering
\scalebox{0.60}{
\begin{tabular}{cc|cccc|ccccc|ccc}
\hline
$t_{\rm start}$$\sim$$t_{\rm stop}$&$S$&$K$&$\alpha$&$E_{\rm c}$&$F$&$K$&$\alpha$&$\beta$&$E_{\rm p}$&$F$&$\Delta$DIC&$p_{\rm DIC,CPL}$&$p_{\rm DIC,Band}$\\
(1)&(2)&(3)&(4)&(5)&(6)&(7)&(8)&(9)&(10)&(11)&(12)&(13)&(14)\\
\hline
$P_{1}$\\
\hline
-0.067$\sim$0.029&8.40&0.37$^{+0.07}_{-0.07}$$\times$10$^{-1}$&-0.98$^{+0.17}_{-0.17}$&891$^{+616}_{-519}$&4.68$^{+4.85}_{-2.45}$$\times$10$^{-6}$&0.37$^{+0.08}_{-0.08}$$\times$10$^{-1}$&-0.98$^{+0.19}_{-0.19}$&-6.17$^{+2.72}_{-2.71}$&921$^{+614}_{-501}$&4.42$^{+4.94}_{-1.99}$$\times$10$^{-6}$&0.0&0.5&0.3\\
0.029$\sim$0.141&21.45&134.00$^{+38.90}_{-38.70}$$\times$10$^{-1}$&-1.11$^{+0.08}_{-0.08}$&1170$^{+517}_{-479}$&10.37$^{+10.65}_{-4.94}$$\times$10$^{-6}$&0.71$^{+0.05}_{-0.05}$$\times$10$^{-1}$&-1.15$^{+0.06}_{-0.07}$&-5.12$^{+2.74}_{-3.12}$&1211$^{+386}_{-370}$&12.53$^{+3.27}_{-3.19}$$\times$10$^{-6}$&81.7&0.6&1.4\\
0.141$\sim$0.294&39.83&1.41$^{+0.08}_{-0.08}$$\times$10$^{-1}$&-1.11$^{+0.05}_{-0.05}$&944$^{+230}_{-232}$&16.99$^{+5.07}_{-3.33}$$\times$10$^{-6}$&1.41$^{+0.08}_{-0.08}$$\times$10$^{-1}$&-1.11$^{+0.05}_{-0.05}$&-6.34$^{+2.50}_{-2.54}$&824$^{+165}_{-165}$&17.33$^{+3.78}_{-2.84}$$\times$10$^{-6}$&0.3&2.6&2.8\\
0.294$\sim$0.415&50.05&2.78$^{+0.17}_{-0.17}$$\times$10$^{-1}$&-0.90$^{+0.05}_{-0.05}$&451$^{+65}_{-66}$&21.90$^{+4.86}_{-3.63}$$\times$10$^{-6}$&2.78$^{+0.18}_{-0.18}$$\times$10$^{-1}$&-0.90$^{+0.05}_{-0.05}$&-5.92$^{+2.68}_{-2.72}$&492$^{+55}_{-56}$&22.58$^{+3.64}_{-3.11}$$\times$10$^{-6}$&0.0&2.8&2.9\\
0.415$\sim$0.546&73.35&4.42$^{+0.18}_{-0.18}$$\times$10$^{-1}$&-0.78$^{+0.04}_{-0.04}$&452$^{+42}_{-42}$&40.15$^{+5.55}_{-5.07}$$\times$10$^{-6}$&4.44$^{+0.18}_{-0.18}$$\times$10$^{-1}$&-0.78$^{+0.04}_{-0.04}$&-6.21$^{+2.53}_{-2.57}$&545$^{+39}_{-40}$&41.10$^{+4.81}_{-3.84}$$\times$10$^{-6}$&-0.0&2.9&3.0\\
0.546$\sim$0.701&96.49&6.47$^{+0.22}_{-0.22}$$\times$10$^{-1}$&-0.66$^{+0.03}_{-0.03}$&354$^{+23}_{-23}$&49.72$^{+5.44}_{-5.02}$$\times$10$^{-6}$&6.63$^{+0.26}_{-0.26}$$\times$10$^{-1}$&-0.65$^{+0.04}_{-0.03}$&-4.05$^{+1.24}_{-1.89}$&454$^{+26}_{-25}$&54.13$^{+6.04}_{-5.79}$$\times$10$^{-6}$&-6.0&2.9&0.8\\
0.701$\sim$1.579&263.62&7.71$^{+0.08}_{-0.08}$$\times$10$^{-1}$&-0.68$^{+0.01}_{-0.01}$&451$^{+10}_{-11}$&80.76$^{+3.02}_{-2.94}$$\times$10$^{-6}$&7.72$^{+0.08}_{-0.08}$$\times$10$^{-1}$&-0.68$^{+0.01}_{-0.01}$&-6.43$^{+1.96}_{-2.31}$&594$^{+10}_{-10}$&81.07$^{+2.16}_{-1.95}$$\times$10$^{-6}$&-1.1&3.0&2.6\\
1.579$\sim$1.713&117.21&8.32$^{+0.16}_{-0.16}$$\times$10$^{-1}$&-0.47$^{+0.02}_{-0.02}$&515$^{+22}_{-22}$&147.00$^{+12.37}_{-11.77}$$\times$10$^{-6}$&8.32$^{+0.16}_{-0.17}$$\times$10$^{-1}$&-0.46$^{+0.02}_{-0.02}$&-6.73$^{+2.01}_{-2.14}$&788$^{+25}_{-25}$&146.90$^{+9.38}_{-8.28}$$\times$10$^{-6}$&-0.1&3.0&3.0\\
1.713$\sim$1.805&88.47&8.93$^{+0.34}_{-0.34}$$\times$10$^{-1}$&-0.56$^{+0.04}_{-0.04}$&319$^{+21}_{-21}$&67.00$^{+8.01}_{-6.55}$$\times$10$^{-6}$&9.12$^{+0.41}_{-0.42}$$\times$10$^{-1}$&-0.54$^{+0.04}_{-0.04}$&-5.38$^{+2.41}_{-2.98}$&447$^{+26}_{-27}$&70.25$^{+10.60}_{-7.62}$$\times$10$^{-6}$&-1.9&2.9&1.9\\
1.805$\sim$1.933&80.82&6.01$^{+0.33}_{-0.33}$$\times$10$^{-1}$&-0.84$^{+0.04}_{-0.04}$&275$^{+26}_{-26}$&28.49$^{+3.60}_{-3.22}$$\times$10$^{-6}$&9.62$^{+1.71}_{-1.71}$$\times$10$^{-1}$&-0.57$^{+0.11}_{-0.11}$&-2.15$^{+0.10}_{-0.09}$&199$^{+27}_{-27}$&42.18$^{+14.55}_{-9.59}$$\times$10$^{-6}$&-20.9&2.9&0.5\\
1.933$\sim$2.137&72.22&3.10$^{+0.16}_{-0.16}$$\times$10$^{-1}$&-1.01$^{+0.04}_{-0.04}$&392$^{+46}_{-46}$&18.98$^{+2.76}_{-2.28}$$\times$10$^{-6}$&3.11$^{+0.16}_{-0.16}$$\times$10$^{-1}$&-1.01$^{+0.04}_{-0.04}$&-6.43$^{+2.35}_{-2.36}$&381$^{+32}_{-32}$&19.17$^{+2.09}_{-1.81}$$\times$10$^{-6}$&0.4&2.9&3.1\\
2.137$\sim$2.406&63.87&2.48$^{+0.16}_{-0.16}$$\times$10$^{-1}$&-1.00$^{+0.05}_{-0.04}$&301$^{+36}_{-36}$&11.75$^{+1.70}_{-1.52}$$\times$10$^{-6}$&2.49$^{+0.17}_{-0.17}$$\times$10$^{-1}$&-1.00$^{+0.05}_{-0.05}$&-6.41$^{+2.39}_{-2.41}$&297$^{+25}_{-26}$&11.82$^{+1.54}_{-1.24}$$\times$10$^{-6}$&0.5&2.8&3.1\\
2.406$\sim$2.452&37.17&2.83$^{+0.19}_{-0.19}$$\times$10$^{-1}$&-1.01$^{+0.06}_{-0.06}$&1040$^{+285}_{-291}$&42.69$^{+18.09}_{-10.96}$$\times$10$^{-6}$&2.90$^{+0.21}_{-0.21}$$\times$10$^{-1}$&-1.00$^{+0.07}_{-0.07}$&-4.30$^{+2.00}_{-3.02}$&916$^{+225}_{-210}$&45.91$^{+15.24}_{-10.74}$$\times$10$^{-6}$&-3.8&2.3&1.0\\
\hline
$P_{2}$\\
\hline
2.452$\sim$2.642&152.49&9.41$^{+0.14}_{-0.14}$$\times$10$^{-1}$&-0.35$^{+0.02}_{-0.02}$&464$^{+15}_{-15}$&169.90$^{+11.71}_{-10.59}$$\times$10$^{-6}$&9.43$^{+0.14}_{-0.14}$$\times$10$^{-1}$&-0.35$^{+0.02}_{-0.02}$&-6.95$^{+1.98}_{-2.08}$&763$^{+17}_{-18}$&170.30$^{+8.70}_{-6.92}$$\times$10$^{-6}$&-0.2&3.0&2.9\\
2.642$\sim$2.882&135.79&6.40$^{+0.10}_{-0.10}$$\times$10$^{-1}$&-0.51$^{+0.02}_{-0.02}$&559$^{+22}_{-22}$&117.20$^{+8.53}_{-7.44}$$\times$10$^{-6}$&6.41$^{+0.10}_{-0.10}$$\times$10$^{-1}$&-0.51$^{+0.02}_{-0.02}$&-5.55$^{+1.68}_{-2.30}$&827$^{+24}_{-23}$&118.90$^{+6.59}_{-5.62}$$\times$10$^{-6}$&-2.4&3.0&2.4\\
2.882$\sim$3.088&102.21&4.47$^{+0.09}_{-0.09}$$\times$10$^{-1}$&-0.53$^{+0.02}_{-0.02}$&666$^{+33}_{-33}$&102.80$^{+9.27}_{-8.80}$$\times$10$^{-6}$&4.47$^{+0.09}_{-0.09}$$\times$10$^{-1}$&-0.53$^{+0.02}_{-0.02}$&-5.93$^{+1.99}_{-2.47}$&974$^{+37}_{-37}$&103.50$^{+6.94}_{-5.88}$$\times$10$^{-6}$&-1.2&3.0&2.5\\
3.088$\sim$3.208&92.58&4.99$^{+0.11}_{-0.11}$$\times$10$^{-1}$&-0.45$^{+0.03}_{-0.03}$&814$^{+50}_{-50}$&180.70$^{+22.92}_{-18.31}$$\times$10$^{-6}$&5.16$^{+0.12}_{-0.12}$$\times$10$^{-1}$&-0.39$^{+0.03}_{-0.03}$&-2.85$^{+0.13}_{-0.13}$&1099$^{+49}_{-49}$&192.00$^{+16.16}_{-15.55}$$\times$10$^{-6}$&-53.8&3.0&4.0\\
3.208$\sim$3.605&147.68&4.58$^{+0.06}_{-0.06}$$\times$10$^{-1}$&-0.35$^{+0.02}_{-0.02}$&619$^{+20}_{-20}$&131.60$^{+10.15}_{-8.08}$$\times$10$^{-6}$&4.64$^{+0.06}_{-0.07}$$\times$10$^{-1}$&-0.33$^{+0.02}_{-0.02}$&-2.88$^{+0.09}_{-0.09}$&968$^{+24}_{-24}$&149.00$^{+7.72}_{-7.26}$$\times$10$^{-6}$&-93.1&3.0&4.0\\
3.605$\sim$3.739&80.43&4.00$^{+0.10}_{-0.10}$$\times$10$^{-1}$&-0.32$^{+0.04}_{-0.04}$&594$^{+36}_{-36}$&115.80$^{+15.62}_{-14.13}$$\times$10$^{-6}$&4.04$^{+0.10}_{-0.10}$$\times$10$^{-1}$&-0.30$^{+0.04}_{-0.04}$&-3.06$^{+0.21}_{-0.20}$&966$^{+42}_{-42}$&129.30$^{+12.41}_{-11.48}$$\times$10$^{-6}$&-19.7&2.9&3.9\\
3.739$\sim$3.959&140.01&6.34$^{+0.10}_{-0.10}$$\times$10$^{-1}$&-0.20$^{+0.02}_{-0.02}$&533$^{+19}_{-19}$&193.50$^{+15.29}_{-14.73}$$\times$10$^{-6}$&6.55$^{+0.11}_{-0.11}$$\times$10$^{-1}$&-0.14$^{+0.03}_{-0.03}$&-2.71$^{+0.07}_{-0.07}$&873$^{+23}_{-23}$&224.70$^{+13.46}_{-13.54}$$\times$10$^{-6}$&-170.4&3.0&4.0\\
3.959$\sim$4.096&129.78&9.68$^{+0.18}_{-0.18}$$\times$10$^{-1}$&-0.19$^{+0.03}_{-0.03}$&399$^{+14}_{-14}$&175.70$^{+14.96}_{-12.87}$$\times$10$^{-6}$&9.75$^{+0.19}_{-0.19}$$\times$10$^{-1}$&-0.19$^{+0.03}_{-0.03}$&-3.65$^{+0.35}_{-0.23}$&709$^{+19}_{-19}$&187.10$^{+12.17}_{-10.94}$$\times$10$^{-6}$&-11.0&3.0&3.8\\
4.096$\sim$4.442&171.47&8.48$^{+0.13}_{-0.13}$$\times$10$^{-1}$&-0.44$^{+0.02}_{-0.02}$&365$^{+11}_{-11}$&90.91$^{+4.75}_{-5.11}$$\times$10$^{-6}$&8.55$^{+0.15}_{-0.15}$$\times$10$^{-1}$&-0.44$^{+0.02}_{-0.02}$&-3.60$^{+0.39}_{-0.21}$&560$^{+13}_{-13}$&97.41$^{+4.96}_{-4.85}$$\times$10$^{-6}$&-9.2&3.0&3.5\\
4.442$\sim$4.509&67.63&7.06$^{+0.35}_{-0.34}$$\times$10$^{-1}$&-0.73$^{+0.04}_{-0.04}$&350$^{+32}_{-33}$&49.84$^{+7.13}_{-6.25}$$\times$10$^{-6}$&7.07$^{+0.35}_{-0.36}$$\times$10$^{-1}$&-0.73$^{+0.05}_{-0.05}$&-5.87$^{+2.69}_{-2.78}$&440$^{+30}_{-30}$&51.28$^{+6.70}_{-5.57}$$\times$10$^{-6}$&-0.4&2.9&2.8\\
4.509$\sim$4.770&142.78&7.16$^{+0.12}_{-0.12}$$\times$10$^{-1}$&-0.65$^{+0.02}_{-0.02}$&493$^{+20}_{-20}$&88.02$^{+5.56}_{-5.44}$$\times$10$^{-6}$&7.16$^{+0.12}_{-0.12}$$\times$10$^{-1}$&-0.65$^{+0.02}_{-0.02}$&-6.78$^{+2.03}_{-2.13}$&666$^{+19}_{-19}$&88.41$^{+4.12}_{-3.73}$$\times$10$^{-6}$&0.0&3.0&3.1\\
4.770$\sim$4.950&134.52&9.70$^{+0.21}_{-0.21}$$\times$10$^{-1}$&-0.45$^{+0.02}_{-0.02}$&351$^{+14}_{-14}$&96.19$^{+6.57}_{-6.45}$$\times$10$^{-6}$&9.70$^{+0.20}_{-0.20}$$\times$10$^{-1}$&-0.45$^{+0.02}_{-0.02}$&-7.33$^{+1.86}_{-1.84}$&542$^{+14}_{-14}$&96.30$^{+4.52}_{-4.29}$$\times$10$^{-6}$&0.6&3.0&3.2\\
4.950$\sim$5.451&184.67&6.94$^{+0.10}_{-0.10}$$\times$10$^{-1}$&-0.62$^{+0.02}_{-0.02}$&422$^{+13}_{-13}$&72.22$^{+3.53}_{-3.51}$$\times$10$^{-6}$&6.94$^{+0.10}_{-0.10}$$\times$10$^{-1}$&-0.62$^{+0.01}_{-0.01}$&-6.97$^{+1.97}_{-2.05}$&582$^{+13}_{-12}$&72.22$^{+2.46}_{-2.36}$$\times$10$^{-6}$&0.1&3.0&3.1\\
5.451$\sim$5.514&77.55&10.10$^{+0.41}_{-0.41}$$\times$10$^{-1}$&-0.42$^{+0.04}_{-0.04}$&302$^{+20}_{-20}$&83.08$^{+10.90}_{-9.34}$$\times$10$^{-6}$&10.12$^{+0.40}_{-0.40}$$\times$10$^{-1}$&-0.41$^{+0.04}_{-0.04}$&-6.81$^{+2.16}_{-2.15}$&475$^{+21}_{-21}$&83.12$^{+6.69}_{-6.74}$$\times$10$^{-6}$&0.5&2.9&3.2\\
5.514$\sim$5.689&109.97&10.10$^{+0.43}_{-0.43}$$\times$10$^{-1}$&-0.55$^{+0.03}_{-0.03}$&193$^{+10}_{-10}$&37.07$^{+3.57}_{-3.20}$$\times$10$^{-6}$&10.49$^{+0.58}_{-0.57}$$\times$10$^{-1}$&-0.53$^{+0.04}_{-0.04}$&-4.13$^{+1.15}_{-1.47}$&270$^{+12}_{-12}$&39.58$^{+4.51}_{-4.22}$$\times$10$^{-6}$&-5.6&2.9&1.1\\
5.689$\sim$5.808&69.83&6.33$^{+0.59}_{-0.59}$$\times$10$^{-1}$&-0.89$^{+0.06}_{-0.06}$&171$^{+20}_{-20}$&17.29$^{+3.08}_{-2.54}$$\times$10$^{-6}$&5.15$^{+0.21}_{-0.21}$$\times$10$^{-1}$&-1.00$^{+0.04}_{-0.04}$&-5.10$^{+2.35}_{-3.08}$&226$^{+5}_{-5}$&19.52$^{+2.96}_{-1.50}$$\times$10$^{-6}$&2.1&2.6&0.9\\
5.808$\sim$6.000&60.35&2.46$^{+0.32}_{-0.32}$$\times$10$^{-1}$&-1.35$^{+0.07}_{-0.08}$&202$^{+41}_{-41}$&8.22$^{+1.64}_{-1.45}$$\times$10$^{-6}$&4.53$^{+1.21}_{-1.15}$$\times$10$^{-1}$&-1.06$^{+0.13}_{-0.13}$&-2.20$^{+0.11}_{-0.11}$&83$^{+11}_{-11}$&11.53$^{+5.05}_{-3.72}$$\times$10$^{-6}$&-18.9&1.6&-0.6\\
\hline
$P_{AO}$\\
\hline
6.000$\sim$6.436&62.10&1.02$^{+0.10}_{-0.11}$$\times$10$^{-1}$&-1.63$^{+0.06}_{-0.06}$&440$^{+133}_{-138}$&5.80$^{+1.17}_{-0.88}$$\times$10$^{-6}$&1.09$^{+0.14}_{-0.16}$$\times$10$^{-1}$&-1.60$^{+0.07}_{-0.08}$&-5.13$^{+2.87}_{-3.31}$&139$^{+29}_{-32}$&5.98$^{+2.08}_{-1.32}$$\times$10$^{-6}$&-2.0&1.0&0.1\\
6.436$\sim$6.867&50.89&0.71$^{+0.07}_{-0.07}$$\times$10$^{-1}$&-1.73$^{+0.05}_{-0.05}$&537$^{+201}_{-182}$&4.66$^{+0.85}_{-0.66}$$\times$10$^{-6}$&0.88$^{+0.14}_{-0.20}$$\times$10$^{-1}$&-1.64$^{+0.10}_{-0.11}$&-3.58$^{+1.50}_{-2.81}$&113$^{+33}_{-40}$&5.21$^{+2.31}_{-1.58}$$\times$10$^{-6}$&-19.6&1.4&-15.2\\
6.867$\sim$8.221&64.64&0.44$^{+0.01}_{-0.01}$$\times$10$^{-1}$&-1.77$^{+0.02}_{-0.02}$&892$^{+84}_{-86}$&3.55$^{+0.14}_{-0.15}$$\times$10$^{-6}$&0.43$^{+0.02}_{-0.05}$$\times$10$^{-1}$&-1.81$^{+-0.00}_{-0.05}$&-4.96$^{+2.68}_{-2.72}$&630$^{+275}_{-265}$&4.75$^{+0.43}_{-0.81}$$\times$10$^{-6}$&-31.8&2.2&-23.1\\
8.221$\sim$9.567&49.76&0.32$^{+0.01}_{-0.01}$$\times$10$^{-1}$&-1.78$^{+0.03}_{-0.03}$&830$^{+132}_{-140}$&2.59$^{+0.14}_{-0.17}$$\times$10$^{-6}$&0.29$^{+0.01}_{-0.01}$$\times$10$^{-1}$&-1.85$^{+0.02}_{-0.03}$&-5.02$^{+2.64}_{-2.64}$&647$^{+264}_{-275}$&3.50$^{+0.22}_{-0.37}$$\times$10$^{-6}$&-6.2&2.2&1.5\\
9.567$\sim$12.400&54.86&0.22$^{+0.01}_{-0.01}$$\times$10$^{-1}$&-1.86$^{+0.03}_{-0.03}$&706$^{+193}_{-192}$&1.83$^{+0.17}_{-0.16}$$\times$10$^{-6}$&0.21$^{+0.02}_{-0.02}$$\times$10$^{-1}$&-1.91$^{+0.06}_{-0.05}$&-5.15$^{+2.52}_{-2.54}$&305$^{+362}_{-221}$&2.33$^{+0.43}_{-0.52}$$\times$10$^{-6}$&-7.6&2.1&-6.5\\
12.400$\sim$15.547&47.53&0.17$^{+0.01}_{-0.01}$$\times$10$^{-1}$&-1.93$^{+0.03}_{-0.03}$&699$^{+207}_{-205}$&1.47$^{+0.14}_{-0.14}$$\times$10$^{-6}$&0.15$^{+0.00}_{-0.00}$$\times$10$^{-1}$&-1.99$^{+0.01}_{-0.01}$&-5.54$^{+2.37}_{-2.31}$&512$^{+321}_{-312}$&2.11$^{+0.06}_{-0.09}$$\times$10$^{-6}$&0.3&2.0&1.0\\
\hline
$P_{3}$\\
\hline
15.547$\sim$15.872&38.28&0.74$^{+0.07}_{-0.07}$$\times$10$^{-1}$&-1.49$^{+0.06}_{-0.06}$&552$^{+199}_{-182}$&4.59$^{+1.03}_{-0.93}$$\times$10$^{-6}$&0.76$^{+0.09}_{-0.10}$$\times$10$^{-1}$&-1.47$^{+0.08}_{-0.08}$&-5.07$^{+2.60}_{-2.62}$&278$^{+76}_{-90}$&4.77$^{+1.65}_{-0.98}$$\times$10$^{-6}$&-1.8&1.7&0.0\\
15.872$\sim$16.173&49.52&1.06$^{+0.09}_{-0.09}$$\times$10$^{-1}$&-1.49$^{+0.05}_{-0.06}$&649$^{+197}_{-214}$&6.94$^{+1.74}_{-1.13}$$\times$10$^{-6}$&1.14$^{+0.08}_{-0.08}$$\times$10$^{-1}$&-1.45$^{+0.05}_{-0.05}$&-6.18$^{+2.61}_{-2.61}$&248$^{+32}_{-31}$&6.59$^{+0.74}_{-0.76}$$\times$10$^{-6}$&0.5&1.7&2.4\\
16.173$\sim$16.927&98.83&1.99$^{+0.14}_{-0.14}$$\times$10$^{-1}$&-1.39$^{+0.04}_{-0.04}$&198$^{+23}_{-23}$&6.71$^{+0.78}_{-0.64}$$\times$10$^{-6}$&2.94$^{+0.38}_{-0.38}$$\times$10$^{-1}$&-1.20$^{+0.07}_{-0.07}$&-2.36$^{+0.09}_{-0.08}$&88$^{+7}_{-7}$&8.23$^{+1.70}_{-1.38}$$\times$10$^{-6}$&-18.7&2.6&3.1\\
16.927$\sim$17.324&59.55&1.19$^{+0.14}_{-0.14}$$\times$10$^{-1}$&-1.59$^{+0.07}_{-0.06}$&245$^{+53}_{-56}$&5.11$^{+0.96}_{-0.77}$$\times$10$^{-6}$&1.23$^{+0.16}_{-0.16}$$\times$10$^{-1}$&-1.57$^{+0.07}_{-0.07}$&-5.79$^{+2.79}_{-2.80}$&96$^{+10}_{-10}$&5.23$^{+1.19}_{-0.93}$$\times$10$^{-6}$&0.7&1.7&2.5\\
17.324$\sim$17.719&51.00&0.86$^{+0.12}_{-0.12}$$\times$10$^{-1}$&-1.69$^{+0.07}_{-0.07}$&283$^{+83}_{-87}$&4.31$^{+0.94}_{-0.74}$$\times$10$^{-6}$&1.00$^{+0.17}_{-0.16}$$\times$10$^{-1}$&-1.62$^{+0.08}_{-0.08}$&-3.45$^{+1.21}_{-2.00}$&70$^{+11}_{-10}$&4.57$^{+1.54}_{-1.06}$$\times$10$^{-6}$&-5.3&-0.1&-1.2\\
17.719$\sim$20.397&95.90&0.64$^{+0.04}_{-0.04}$$\times$10$^{-1}$&-1.72$^{+0.04}_{-0.03}$&210$^{+26}_{-27}$&3.04$^{+0.29}_{-0.28}$$\times$10$^{-6}$&0.77$^{+0.10}_{-0.10}$$\times$10$^{-1}$&-1.63$^{+0.06}_{-0.06}$&-2.86$^{+0.48}_{-0.09}$&53$^{+4}_{-4}$&3.27$^{+0.77}_{-0.59}$$\times$10$^{-6}$&-14.2&2.5&-3.3\\
20.397$\sim$21.699&50.58&0.34$^{+0.03}_{-0.03}$$\times$10$^{-1}$&-1.95$^{+0.04}_{-0.04}$&197$^{+36}_{-37}$&2.32$^{+0.29}_{-0.26}$$\times$10$^{-6}$&0.64$^{+0.13}_{-0.12}$$\times$10$^{-1}$&-1.65$^{+0.08}_{-0.08}$&-3.21$^{+0.67}_{-0.11}$&31$^{+1}_{-1}$&2.11$^{+0.68}_{-0.55}$$\times$10$^{-6}$&6.0&1.7&-2.4\\
21.699$\sim$23.330&37.70&0.21$^{+0.02}_{-0.02}$$\times$10$^{-1}$&-1.96$^{+0.04}_{-0.04}$&240$^{+51}_{-53}$&1.49$^{+0.18}_{-0.16}$$\times$10$^{-6}$&0.27$^{+0.06}_{-0.06}$$\times$10$^{-1}$&-1.85$^{+0.09}_{-0.08}$&-4.14$^{+1.83}_{-2.89}$&17$^{+5}_{-5}$&1.55$^{+0.71}_{-0.62}$$\times$10$^{-6}$&-8.1&1.7&-7.0\\
23.330$\sim$26.530&39.21&0.14$^{+0.02}_{-0.02}$$\times$10$^{-1}$&-1.91$^{+0.06}_{-0.06}$&347$^{+122}_{-113}$&1.04$^{+0.18}_{-0.14}$$\times$10$^{-6}$&0.19$^{+0.05}_{-0.05}$$\times$10$^{-1}$&-1.82$^{+0.12}_{-0.10}$&-4.78$^{+2.52}_{-2.73}$&31$^{+8}_{-8}$&1.04$^{+0.52}_{-0.39}$$\times$10$^{-6}$&-6.9&1.2&-5.5\\
26.530$\sim$33.075&37.67&0.10$^{+0.01}_{-0.01}$$\times$10$^{-1}$&-1.93$^{+0.05}_{-0.05}$&387$^{+123}_{-116}$&0.75$^{+0.10}_{-0.09}$$\times$10$^{-6}$&0.11$^{+0.01}_{-0.02}$$\times$10$^{-1}$&-1.88$^{+0.06}_{-0.07}$&-5.01$^{+2.57}_{-2.69}$&30$^{+8}_{-9}$&0.75$^{+0.28}_{-0.21}$$\times$10$^{-6}$&0.4&1.5&1.5\\
33.075$\sim$47.327&35.24&0.07$^{+0.00}_{-0.00}$$\times$10$^{-1}$&-1.84$^{+0.03}_{-0.03}$&493$^{+81}_{-83}$&0.53$^{+0.04}_{-0.04}$$\times$10$^{-6}$&0.07$^{+0.01}_{-0.01}$$\times$10$^{-1}$&-1.87$^{+0.08}_{-0.10}$&-4.59$^{+2.50}_{-2.80}$&74$^{+20}_{-27}$&0.62$^{+0.28}_{-0.19}$$\times$10$^{-6}$&-11.8&2.2&-7.3\\
47.327$\sim$73.490&28.11&0.05$^{+0.01}_{-0.01}$$\times$10$^{-1}$&-1.83$^{+0.06}_{-0.06}$&583$^{+268}_{-250}$&0.35$^{+0.06}_{-0.06}$$\times$10$^{-6}$&0.06$^{+0.01}_{-0.01}$$\times$10$^{-1}$&-1.71$^{+0.08}_{-0.07}$&-4.03$^{+1.90}_{-2.91}$&67$^{+15}_{-14}$&0.34$^{+0.12}_{-0.09}$$\times$10$^{-6}$&-2.7&1.0&-0.7\\
\hline
\end{tabular}
}
\caption{Time-resolved spectral fit results of GRB 190114C. (1) The start and stop times (in units of s) of the BBlocks time bins; (2) the significance $S$; (3-6) the best-fit parameters for the CPL model; (7-11) the best-fit parameters for the Band model; (12) the difference between the Deviance Information Criterion (DIC) for the CPL and the Band model, $\Delta$DIC=DIC$_{\rm Band}$-DIC$_{\rm CPL}$; (13-14) the effective number of parameters ($p_{\rm DIC}$) for the CPL and Band model, respectively.}\label{tab:190114C}
\end{table}

\clearpage
\begin{table}
\setlength{\tabcolsep}{0.15em}
\renewcommand\arraystretch{1.2}
\centering
\scalebox{0.80}{
\begin{tabular}{c|cccccccc}
\hline
&$t_{\rm start} \sim t_{\rm stop}$ &$S$&Model&$\Delta$DIC&Temperature&Thermal Flux&Total Flux&Ratio\\
&(s)&&&DIC$_{\rm (CPL+BB)-(CPL)}$&(keV)&(erg cm$^{-2}$ s$^{-1}$)&(erg cm$^{-2}$ s$^{-1})$ \\
\hline
&0.55$\sim$0.70&96.49&CPL+BB&-177&351$^{+99}_{-100}$&0.59$^{+1.82}_{-0.45}$$\times$10$^{-5}$&0.57$^{+0.19}_{-0.11}$$\times$10$^{-4}$&0.10$^{+0.32}_{-0.08}$\\
&0.70$\sim$0.98&153.65&CPL+BB&-285&283$^{+59}_{-55}$&0.58$^{+1.27}_{-0.42}$$\times$10$^{-5}$&0.72$^{+0.15}_{-0.11}$$\times$10$^{-4}$&0.08$^{+0.18}_{-0.06}$\\
&0.98$\sim$1.45&196.39&CPL+BB&-43&186$^{+40}_{-34}$&0.91$^{+1.72}_{-0.80}$$\times$10$^{-5}$&0.86$^{+0.22}_{-0.24}$$\times$10$^{-4}$&0.10$^{+0.20}_{-0.10}$\\
$P^{\rm th}_{1}$&1.45$\sim$1.58&105.17&CPL+BB&-148&163$^{+20}_{-19}$&2.42$^{+2.10}_{-1.13}$$\times$10$^{-5}$&1.05$^{+0.03}_{-0.20}$$\times$10$^{-4}$&0.23$^{+0.20}_{-0.11}$\\
&1.58$\sim$1.64&80.78&CPL+BB&-29&151$^{+11}_{-11}$&5.26$^{+2.68}_{-1.69}$$\times$10$^{-5}$&1.38$^{+0.33}_{-0.28}$$\times$10$^{-4}$&0.36$^{+0.21}_{-0.14}$\\
&1.64$\sim$1.71&83.47&CPL+BB&-21&136$^{+28}_{-24}$&2.06$^{+4.18}_{-1.63}$$\times$10$^{-5}$&1.49$^{+0.68}_{-0.41}$$\times$10$^{-4}$&0.14$^{+0.29}_{-0.12}$\\
&1.71$\sim$1.80&88.47&CPL+BB&-114&73$^{+13}_{-18}$&0.09$^{+0.20}_{-0.07}$$\times$10$^{-5}$&0.71$^{+0.17}_{-0.14}$$\times$10$^{-4}$&0.01$^{+0.03}_{-0.01}$\\
&1.80$\sim$1.93&80.82&CPL+BB&-139&30$^{+4}_{-5}$&0.07$^{+0.05}_{-0.04}$$\times$10$^{-5}$&0.32$^{+0.02}_{-0.02}$$\times$10$^{-4}$&0.02$^{+0.02}_{-0.01}$\\
\hline
&2.45$\sim$2.64&152.16&CPL+BB&-21&174$^{+13}_{-13}$&3.88$^{+1.89}_{-1.40}$$\times$10$^{-5}$&1.76$^{+0.26}_{-0.24}$$\times$10$^{-4}$&0.22$^{+0.11}_{-0.09}$\\
&2.64$\sim$2.88&136.19&CPL+BB&-622&197$^{+17}_{-16}$&4.07$^{+1.91}_{-1.34}$$\times$10$^{-5}$&1.17$^{+0.28}_{-0.28}$$\times$10$^{-4}$&0.35$^{+0.18}_{-0.14}$\\
&2.88$\sim$3.09&103.16&CPL+BB&-12&188$^{+22}_{-22}$&2.36$^{+1.83}_{-1.11}$$\times$10$^{-5}$&1.01$^{+0.38}_{-0.29}$$\times$10$^{-4}$&0.23$^{+0.20}_{-0.13}$\\
&3.09$\sim$3.21&92.86&CPL+BB&-20&162$^{+12}_{-8}$&4.93$^{+1.71}_{-1.58}$$\times$10$^{-5}$&1.95$^{+0.85}_{-0.57}$$\times$10$^{-4}$&0.25$^{+0.14}_{-0.11}$\\
&3.21$\sim$3.60&146.18&CPL+BB&-37&149$^{+7}_{-8}$&3.16$^{+1.03}_{-0.79}$$\times$10$^{-5}$&1.36$^{+0.47}_{-0.31}$$\times$10$^{-4}$&0.23$^{+0.11}_{-0.08}$\\
&3.60$\sim$3.74&82.69&CPL+BB&-20&151$^{+11}_{-11}$&3.36$^{+1.65}_{-1.28}$$\times$10$^{-5}$&1.23$^{+0.84}_{-0.49}$$\times$10$^{-4}$&0.27$^{+0.23}_{-0.15}$\\
&3.74$\sim$3.96&140.29&CPL+BB&-63&140$^{+5}_{-5}$&7.11$^{+1.13}_{-1.01}$$\times$10$^{-5}$&2.23$^{+0.81}_{-0.55}$$\times$10$^{-4}$&0.32$^{+0.13}_{-0.09}$\\
$P^{\rm th}_{2}$&3.96$\sim$4.10&130.66&CPL+BB&-44&108$^{+6}_{-7}$&3.52$^{+1.85}_{-1.83}$$\times$10$^{-5}$&1.79$^{+1.71}_{-0.79}$$\times$10$^{-4}$&0.20$^{+0.22}_{-0.13}$\\
&4.10$\sim$4.44&170.04&CPL+BB&-922&115$^{+7}_{-7}$&1.89$^{+0.80}_{-0.59}$$\times$10$^{-5}$&0.94$^{+0.13}_{-0.11}$$\times$10$^{-4}$&0.20$^{+0.09}_{-0.07}$\\
&4.44$\sim$4.51&69.23&CPL+BB&-207&97$^{+15}_{-13}$&1.22$^{+1.57}_{-0.74}$$\times$10$^{-5}$&0.55$^{+0.23}_{-0.13}$$\times$10$^{-4}$&0.22$^{+0.30}_{-0.15}$\\
&4.51$\sim$4.77&142.52&CPL+BB&-133&111$^{+5}_{-5}$&3.45$^{+1.02}_{-0.72}$$\times$10$^{-5}$&0.85$^{+0.22}_{-0.16}$$\times$10$^{-4}$&0.41$^{+0.16}_{-0.12}$\\
&4.77$\sim$4.95&134.52&CPL+BB&-54&90$^{+4}_{-4}$&3.21$^{+0.89}_{-0.76}$$\times$10$^{-5}$&0.97$^{+0.33}_{-0.23}$$\times$10$^{-4}$&0.33$^{+0.15}_{-0.11}$\\
&4.95$\sim$5.45&184.46&CPL+BB&-176&91$^{+3}_{-3}$&2.45$^{+0.45}_{-0.39}$$\times$10$^{-5}$&0.70$^{+0.14}_{-0.10}$$\times$10$^{-4}$&0.35$^{+0.09}_{-0.07}$\\
&5.45$\sim$5.51&76.02&CPL+BB&-93&82$^{+5}_{-5}$&3.18$^{+1.30}_{-1.22}$$\times$10$^{-5}$&0.80$^{+0.95}_{-0.32}$$\times$10$^{-4}$&0.40$^{+0.49}_{-0.22}$\\
&5.51$\sim$5.65&100.84&CPL+BB&-27&53$^{+4}_{-4}$&1.10$^{+0.51}_{-0.43}$$\times$10$^{-5}$&0.44$^{+0.11}_{-0.16}$$\times$10$^{-4}$&0.25$^{+0.13}_{-0.12}$\\
&5.65$\sim$5.69&48.93&CPL+BB&-26&32$^{+2}_{-2}$&1.07$^{+0.33}_{-0.26}$$\times$10$^{-5}$&0.35$^{+0.07}_{-0.06}$$\times$10$^{-4}$&0.30$^{+0.11}_{-0.09}$\\
\hline
$T_{90}$&0.00$\sim$116.00&190.61&CPL+BB&-266&132$^{+4}_{-4}$&0.12$^{+0.04}_{-0.03}$$\times$10$^{-5}$&0.06$^{+0.01}_{-0.01}$$\times$10$^{-4}$&0.21$^{+0.07}_{-0.05}$\\
\hline
\end{tabular}
}
\caption{Spectral parameters of the slices having a thermal component in GRB 190114C. The spectra are best fitted by a two-component scenario, with a thermal BB component accompanied by a non-thermal CPL component. The table lists the start and stops times of the BBlocks slices, the significance, the best-fitted model, the $\Delta$DIC between CPL+BB and CPL models, the temperature, the thermal and total flux, and the ratio of thermal flux. Flux is defined in the energy band of $1$~keV to $10$~MeV. For the slices of $\sim 3$~s to $\sim 4$~s, Band+BB offers very close goodness of fitting as CPL+BB, for the global consistency, and considering the time-integrated spectrum is best fitted by CPL+BB, here we perform all the thermal analysis using CPL+BB.}
\label{tab:CPLvsCPLBB}
\end{table}

\clearpage
\begin{table*}
\setlength{\tabcolsep}{0.10em}
\renewcommand\arraystretch{1.2}
\caption{Thermal-pulse properties of GRB 190114C}\label{tab:Pulse}
\centering
\scalebox{0.80}{
\begin{tabular}{c|ccc}
\hline
&$P^{\rm th}_{1}$ &$P^{\rm th}_{2}$ \\
&(From $t_{\rm obs}=$0.55~s to 1.93~s)&(From $t_{\rm obs}=$2.45~s to 5.69~s)\\
\hline
Observed properties\\
\hline
Duration&1.38~s&3.24~s\\
Spectral cut-off energy [$E_{\rm c}$]&337$^{+27}_{-27}$(keV)&605$^{+17}_{-17}$(keV)\\
Temperature [k$T$] &267$^{+22}_{-18}$(keV)&145$^{+3}_{-3}$(keV)\\
Thermal energy flux [$F_{\rm BB}$]&(1.51$^{+0.97}_{-0.68}$)$\times$10$^{-5}$(erg~cm$^{-2}$~s$^{-1}$)&(2.32$^{+0.35}_{-0.30}$)$\times$10$^{-5}$(erg~cm$^{-2}$~s$^{-1}$)\\
Total energy flux [$F_{\rm tot}$]&(8.65$^{+1.64}_{-1.34}$)$\times$10$^{-5}$(erg~cm$^{-2}$~s$^{-1}$)&(1.07$^{+0.05}_{-0.05}$)$\times$10$^{-4}$(erg~cm$^{-2}$~s$^{-1}$)\\
Flux ratio [$F_{\rm BB}/F_{\rm tot}$]&0.17$^{+0.12}_{-0.08}$&0.22$^{+0.03}_{-0.03}$\\
Thermal fluence [$S_{\rm BB}$]&(2.09$^{+1.35}_{-0.94}$)$\times$10$^{-5}$(erg~cm$^{-2}$)&(7.51$^{+1.13}_{-0.97}$)$\times$10$^{-5}$(erg~cm$^{-2}$)\\
Total fluence [$S_{\rm tot}$]&(1.20$^{+0.23}_{-0.19}$)$\times$10$^{-4}$(erg~cm$^{-2}$)&(3.46$^{+0.17}_{-0.17}$)$\times$10$^{-4}$(erg~cm$^{-2}$)\\
Isotropic thermal luminosity [$L_{\rm BB, \gamma, iso}$]&(1.04$^{+0.67}_{-0.47}$)$\times$10$^{52}$(erg~s$^{-1}$)&(1.60$^{+0.24}_{-0.21}$)$\times$10$^{52}$(erg~s$^{-1}$)\\
Isotropic total luminosity [$L_{\rm \gamma, iso}$]&(5.95$^{+1.13}_{-0.92}$)$\times$10$^{52}$(erg~s$^{-1}$)&(7.36$^{+0.34}_{-0.34}$)$\times$10$^{52}$(erg~s$^{-1}$)\\
Isotropic thermal energy [$E_{\rm BB, \gamma, iso}$]&(1.01$^{+0.65}_{-0.46}$)$\times$10$^{52}$(erg)&(3.64$^{+0.55}_{-0.47}$)$\times$10$^{52}$(erg)\\
Isotropic total energy [$E_{\rm \gamma, iso}$]&(5.81$^{+1.10}_{-0.91}$)$\times$10$^{52}$(erg)&(1.68$^{+0.08}_{-0.08}$)$\times$10$^{53}$(erg)\\
\hline
Photospheric properties\\
\hline
Nozzle radius [$r_{\rm 0}$]&(8.55$\pm$2.8)$\times$10$^{6}$(cm)&(5.00$\pm$0.48)$\times$10$^{7}$(cm)\\
Saturation radius [$r_{\rm s}$]&(4.31$\pm$1.53)$\times$10$^{9}$(cm)&(1.96$\pm$0.19)$\times$10$^{10}$(cm)\\
Photospheric radius [$r_{\rm ph}$]&(5.33$\pm$0.47)$\times$10$^{11}$(cm)&(1.41$\pm$0.04)$\times$10$^{12}$(cm)\\
\hline
Parameter evolution\\
\hline
Temperature [k$T$(t)] &$\propto t^{-0.93\pm0.04}$&$\propto t^{-1.32\pm0.09}$\\
Effective transverse size [$\Re$(t)]&$\propto t^{3.12\pm0.49}$&$\propto t^{2.37\pm0.32}$\\
Bulk Lorentz factor [$\Gamma$(t)]&$\propto t^{-0.48\pm0.05}$&$\propto t^{-0.81\pm0.08}$\\
Nozzle radius $R_{\rm 0}$ (cm)&$\propto t^{4.69\pm3.89}$&$\propto t^{4.10\pm0.86}$\\
Saturation radius $R_{\rm s}$ (cm)&$\propto t^{4.47\pm4.20}$&$\propto t^{2.10\pm0.96}$\\
photospheric radius $R_{\rm ph}$ (cm)&$\propto t^{1.76\pm0.60}$&$\propto t^{1.09\pm0.37}$\\
\hline
\end{tabular}
}
\end{table*}

\clearpage
\begin{table*}[ht!]
\setlength{\tabcolsep}{-0.5em}
\renewcommand\arraystretch{1.2}
\caption{Global properties of GRB 190114C}\label{tab:global}
\centering
\scalebox{0.80}{
\begin{tabular}{c|c}
\hline
\hline
Measured Parameters\\
\hline
Isotropic equivalent thermal energy [$E_{\rm th, iso}$]&(3.69$^{+0.78}_{-0.67}$)$\times$10$^{52}$~erg\\
Isotropic equivalent non-thermal energy [$E_{\rm nth, iso}$]&[(1.92$^{+0.12}_{-0.11}$)$\times$10$^{53}$(GBM)+8.49$^{+1.80}_{-1.80}$$\times$10$^{51}$(LAT)]~erg\\
Thermal energy flux [$F^{\rm obs}_{\rm BB}$] &(1.27$^{+0.27}_{-0.23}$) $\times$ 10$^{-5}$ ~erg~cm$^{-2}$s$^{-1}$\\
Total energy flux [$F^{\rm obs}_{\gamma}$] &[(7.91$^{+0.33}_{-0.32}$)$\times$10$^{-5}$(GBM)+3.41$^{+0.69}_{-0.69}$$\times$10$^{-6}$(LAT)] ~erg~cm$^{-2}s^{-1}$\\
Deceleration time [$t_{\rm dec}$]&6-10~s\\
Temperature [$kT^{\rm obs}$]&163$\pm$6 ~keV\\
Redshift [$z$] & 0.4254$\pm$0.0005\\
\hline
Derived Parameters\\
\hline
Dimensionless specific enthalpy [$\eta$]&$854\pm38$\\
Bulk Lorentz factor at $r_{\rm ph}$ [$\Gamma_{\rm ph}$]&$833\pm38$\\
Initial Lorentz factor [$\Gamma_{\rm s}$]&$719\pm59$\\
Isotropic equivalent total mass [$M_{\rm iso}$]&$(1.7\pm0.4)\times10^{-3} ~M_\odot$\\
Isotropic kinetic energy [$E_{\rm k, iso}$]& $(1.6\pm0.7)\times10^{54}$ ~erg\\
Isotropic total energy [$E_{\rm tot, iso}$] &$(1.8\pm0.7)\times10^{54}$ ~erg\\
$\gamma$-ray radiative efficiency [$\eta_{\gamma}$]&$15.8\pm5.4~\%$\\
\hline
Further Derived Parameters\\
\hline
Energy fractions assigned to electric fields [$\epsilon_{e,-1}$]&1.11$\pm$0.01\\
Energy fractions assigned to magnetic fields [$\epsilon_{\rm B,-2}$]&0.05$\pm$0.01\\
Characteristic synchrotron frequency [$\nu_{\rm m}$]&(1.30$\pm$0.82)$\times$10$^{17}$ Hz\\
Cooling frequency [$\nu_{\rm c}$]&(4.44$\pm$0.66)$\times$ 10$^{17}$ Hz\\
Klein-Nishina frequency [$\nu_{\rm KN}$]&(6.55$\pm$0.16)$\times$10$^{17}$ Hz\\
\hline
\end{tabular}
}
\end{table*}

\clearpage

\clearpage

\clearpage

\clearpage

\clearpage

\clearpage

\clearpage

\clearpage

\end{document}